%% file: main.tex
\documentclass[aps,pra,reprint,superscriptaddress,nofootinbib,floatfix]{revtex4-2}

\usepackage{amsmath,amssymb,bm}
\usepackage{graphicx}
\usepackage{booktabs}
\usepackage{braket}
\usepackage{xcolor}
\usepackage{microtype}
\usepackage{hyperref}
\hypersetup{
    colorlinks=true,
    citecolor=blue,
    urlcolor=blue,
    linkcolor=blue}

\begin{document}

\title{Distinct Signatures of Dynamical and Spectral Criticality in Quenched Quantum Batteries}

\author{Zheng Liu}
\affiliation{School of Physics, Dalian University of Technology, Dalian 116024, P.R. China}
\author{Wen-Hui Nie}
\affiliation{School of Physics, Dalian University of Technology, Dalian 116024, P.R. China}
\author{Yi-jia Yang}
\affiliation{School of Physics, Dalian University of Technology, Dalian 116024, P.R. China}
\author{Lin-Cheng Wang}
\affiliation{School of Physics, Dalian University of Technology, Dalian 116024, P.R. China}
\author{Chang-shui Yu}
\email{ycs@dlut.edu.cn}
\affiliation{School of Physics, Dalian University of Technology, Dalian 116024, P.R. China}

\begin{abstract}
Criticality is widely used to enhance energy storage in quantum batteries. Two criticalities are compared here, one spectral and one dynamical. We ask two questions. Where does the enhancement that criticality brings to a quenched battery come from? And does the best charging performance require a critical point or a critical time? In the anisotropic XY chain, whose dynamics decomposes into independent momentum sectors, the weight each sector transfers to its upper level organizes the response. A closing spectral gap reshapes that weight near the closing momentum, so the stored energy responds sharply and controllably to the quench parameters. What it supplies is tunability rather than a larger maximum. A dynamical quantum phase transition (DQPT) marks the time at which the evolving state becomes orthogonal to its initial state, where the stored energy stays smooth, and equals the energy maximum when the final precession frequencies coincide, as for a flat band away from the critical field. The interacting Ising chain has no momentum sectors, yet its energy maximum and its return peak separate by a tunable margin. Weak quenches reduce the stored energy to the static susceptibility of the quenched operator. These results identify what optimization acts on, the distribution of transition weights and the spread of precession frequencies, and turn the weak-quench response into a probe of the transition order.
\end{abstract}

\maketitle

\section{Introduction}

A quantum battery stores energy by driving a many-body system out of equilibrium \cite{Campaioli2018,RevModPhys.96.031001,Ferraro2026}, and its performance is usually judged by the stored or extractable energy, the charging power, and their fluctuations. How much energy a quench deposits depends on the initial and final Hamiltonians together, through their spectra and their eigenstates, and this dependence has made criticality a focus of charging studies. Early work showed that entangling operations can raise the extractable work \cite{PhysRevE.87.042123} and that collective charging can enhance the power beyond the sum of the individual contributions \cite{Binder2015,Campaioli2017}. Later studies examined collective solid-state charging \cite{Ferraro2018}, transmon-resonator implementations \cite{PhysRevA.107.023725}, dissipative and charger-mediated open-system charging \cite{PhysRevLett.122.210601,PhysRevB.99.035421}, correlations that suppress extractable work \cite{PhysRevLett.122.047702}, non-Hermitian charging protocols \cite{PhysRevA.109.042207}, self-discharge \cite{d9k1-75d4}, the role of charging fluctuations in establishing a reliable quantum advantage \cite{6kwv-z6fx}, and bounds on the charging speedup imposed by the locality of the charging operations \cite{PhysRevLett.128.140501}. Many-body interactions can modify the charging power, but the mechanism and even the presence of an advantage are model dependent \cite{Campaioli2017,Ferraro2018,Andolina2019PRB}. At equilibrium a continuous quantum phase transition is characterized by a singular ground-state response and by diverging correlation lengths, often accompanied by a closing low-energy gap \cite{Sachdev2011,RevModPhys.69.315,MatthiasVojta_2003}, and topological phases are classified by bulk invariants together with the symmetries relevant to the classification \cite{RevModPhys.82.3045,RevModPhys.83.1057,Ryu_2010,RevModPhys.88.035005}, as the original Su--Schrieffer--Heeger work first described for solitons in long-chain polyenes \cite{SSH}. Quantum-battery studies have since considered interacting spin chains \cite{PhysRevA.97.022106}, topological photonic environments \cite{PhysRevLett.134.180401}, and many-body-localized regimes \cite{PhysRevB.100.115142}, and the quantum-state advantage in charging is not an entanglement monotone, since highly entangled states can help or hinder depending on the protocol \cite{10.1116/5.0184903}. These works probe different aspects of charging dynamics, and to our knowledge none of them establishes a general relation between a phase-transition signature and a charging optimum.

Following a sudden quench, a dynamical quantum phase transition (DQPT) is conventionally identified by a nonanalyticity of the thermodynamic Loschmidt return rate at a real critical time \cite{Heyl2013,Heyl2018,Heyl_2019}, and crossing an equilibrium critical line is in general neither necessary nor sufficient for one \cite{PhysRevB.89.161105,srx7-cpl4}. Topological and geometric descriptions of DQPTs have been established \cite{PhysRevB.91.155127,PhysRevB.93.144306,PhysRevLett.121.130603}, and earlier quench studies used the Bose--Hubbard model to follow fluctuations and correlations across the superfluid--Mott transition \cite{PhysRevLett.97.200601,PhysRevA.77.043615} and spinor and $O(N)$ systems to study topological-defect production and its scaling \cite{PhysRevLett.99.120407,PhysRevD.81.025017,Uhlmann_2010}. The later literature has covered Floquet driving \cite{PhysRevB.100.085308}, finite-time linear ramps \cite{PhysRevA.110.042209}, noisy long-range ramps \cite{PhysRevB.110.064302}, flux quenches \cite{NIE2024130110}, and extended SSH models \cite{PhysRevB.110.054312}. DQPT signatures have been observed in trapped-ion Ising systems \cite{Jurcevic2017}, in topological quenches of ultracold atoms with time- and momentum-resolved tomography \cite{Flaschner2018}, in superconducting-qubit simulations \cite{Guo2019}, in photonic quantum walks \cite{PhysRevLett.122.020501}, and in spinor condensates \cite{PhysRevLett.124.043001}, and the dynamics has been simulated on a trapped-ion quantum processor \cite{Gover2026}, while related trapped-ion experiments have observed another form of many-body dynamical phase transition in large Ising simulators \cite{Zhang2017}. The momentum-resolved atomic measurements are especially relevant here, because they tie a singular return feature to one particular critical momentum, whereas the stored battery energy integrates over momentum sectors with energy-dependent weights. Recent studies of integrable and free-fermion quantum batteries have reported a strong and occasionally nonanalytic dependence of the stored energy on the equilibrium phase structure \cite{PhysRevLett.133.197001,GRAZI2025116383,en18236116}, including a universal analysis based on a Dirac-cone model and checked in Ising and Haldane lattice models \cite{en18236116} and enhanced energy and power near DQPT critical times for appropriate quenches in a topological battery \cite{42rw-l31q}. These works establish that criticality matters but leave open where its contribution comes from and whether it determines the best performance.

We therefore ask two questions. Where does the enhancement that criticality brings to a quenched battery originate? And is the best charging performance tied to a critical point or to a critical time? The two questions are of different kinds, since a spectral transition concerns the closing of an excitation gap whereas a DQPT concerns the return properties of the evolving many-body state. The comparison is confined to these two criticalities. We answer them for a battery prepared in the ground state of an initial Hamiltonian $H_i$ and charged by a sudden change to $H_f$, with the stored energy measured by $H_i$. Three control parameters are available, the preparation, the final Hamiltonian, and the extraction time. The analytical part concerns two-band free fermions with independent momentum sectors, illustrated by the anisotropic XY chain, where the overlap between the initial and final Bloch directions of a sector fixes the largest population that the sector can transfer, and it is supplemented by exact diagonalization of an interacting next-nearest-neighbor Ising chain, in which that decomposition is lost. What emerges is a separation between the part of criticality that can be tuned and the part that cannot, together with a weak-quench limit, taken at fixed size and small quench amplitude, in which the charging response becomes an equilibrium static susceptibility and also reports the order of the underlying phase transition.

The remainder of this paper is organized as follows. Section~\ref{sec:framework} defines the quench-charging protocol and the quantity to be optimized, Sec.~\ref{sec:two_band} develops the analytical framework for two-band free fermions, Sec.~\ref{sec:results} introduces the XY and interacting Ising models where they are used and presents the results around the three control parameters, and Sec.~\ref{sec:conclusion} collects the design rules. Detailed derivations and numerical procedures are given in the appendices.

\section{Quench charging protocol}
\label{sec:framework}

We consider a battery prepared in the ground state $|\psi_0\rangle$ of an initial Hamiltonian $H_i$, whose ground-state energy we denote by $E_0^i$. At time $t=0$ the Hamiltonian is suddenly changed to $H_f$, and the state evolves as $|\psi(t)\rangle=e^{-iH_ft}|\psi_0\rangle$, where we set $\hbar=1$ throughout. The stored energy is the excess energy retained in the initial spectrum,
\begin{equation}
\Delta E(t)=\langle\psi(t)|H_i|\psi(t)\rangle-E_0^i.
\label{eq:stored_energy}
\end{equation}
Here $H_f$ generates the charging dynamics, while $H_i$ defines the energy readout. Thus $\Delta E(t)$ is the energy that survives in the initial spectrum when the charging Hamiltonian is removed at time $t$. The expectation value of $H_f$ is conserved during the evolution, so the charging process is visible as the change of $\langle H_i\rangle$. For a battery of $N$ sites we write the stored-energy density as $\Delta e_N(t)=\Delta E(t)/N$. The optimization problem we address is to make this density as large as possible at the extraction time of our choice, and to identify which properties of the initial and final Hamiltonians control that maximum.

Two further quantities enter the comparison, one for each kind of criticality. The dynamical side is carried by the Loschmidt amplitude $\mathcal G(t)=\langle\psi_0|\psi(t)\rangle$ and the finite-size return rate $\lambda_N(t)=-N^{-1}\ln|\mathcal G(t)|^2$, where the prefactor $N^{-1}$ makes the rate intensive \cite{Heyl2013}. A DQPT corresponds to a nonanalyticity of this rate in the thermodynamic limit, and for the finite interacting chains we characterize the return dynamics by the maxima of $\lambda_N(t)$. The spectral side is carried by the equilibrium excitation gap, the energy difference between the ground state of a Hamiltonian and its first excited state. At a continuous equilibrium transition this gap closes and the ground-state response to the control parameter becomes singular, while a first-order transition leaves the gap finite. In the translationally invariant models studied below the lowest excitation carries a definite momentum, so the gap is a band separation that vanishes at a single closing momentum $k_0$, and the singular response is confined to a neighborhood of that momentum \cite{Sachdev2011,RevModPhys.69.315,MatthiasVojta_2003}. The two sides are logically independent, since the gap belongs to the spectrum of a single Hamiltonian whereas the return rate belongs to the evolution that the quench sets in motion, and they act on the stored energy in different ways \cite{srx7-cpl4}.

\section{Charging dynamics in two-band free-fermion systems}
\label{sec:two_band}

We now restrict the analysis to translationally invariant two-band free-fermion models whose dynamics factorizes into independent two-level momentum sectors, including number-conserving forms and the Bogoliubov--de Gennes forms that arise from spin chains through a Jordan--Wigner transformation \cite{Jordan1928,PhysRevA.2.1075}. The results of this section therefore apply to this free-fermion class rather than to arbitrary interacting many-body batteries, and the interacting chain is treated separately by exact diagonalization. After removing a scalar energy offset in each sector, the Hamiltonian takes the form
\begin{equation}
h_\alpha(k)=\mathbf d^\alpha(k)\cdot\bm\sigma .
\end{equation}
Here $\alpha=i,f$ labels the initial and final Hamiltonian, $k$ is the lattice momentum, $\mathbf d^\alpha(k)$ is the Bloch vector that fixes the orientation and scale of the sector Hamiltonian, and $\bm\sigma=(\sigma_x,\sigma_y,\sigma_z)$ collects the three Pauli matrices. We write $\epsilon_\alpha(k)=|\mathbf d^\alpha(k)|$ for the positive band energy, so that the two levels of the sector are $\pm\epsilon_\alpha(k)$, and $\hat{\mathbf d}^{\,\alpha}(k)=\mathbf d^\alpha(k)/\epsilon_\alpha(k)$ for the corresponding unit Bloch vector.

We write $|\psi_{0,k}\rangle$ for the initial state of a sector, the lower eigenstate of $h_i(k)$, whose Bloch vector is $-\hat{\mathbf d}^{\,i}(k)$. The return amplitude of the sector is the overlap of this state with its evolution, $\mathcal G_k(t)=\langle\psi_{0,k}|e^{-ih_f(k)t}|\psi_{0,k}\rangle$. Writing $m_k=\hat{\mathbf d}^{\,i}(k)\cdot\hat{\mathbf d}^{\,f}(k)$ for the overlap of the two Hamiltonian directions, we obtain
\begin{equation}
\mathcal G_k(t)=\cos[\epsilon_f(k)t]+i m_k\sin[\epsilon_f(k)t].
\label{eq:Gk}
\end{equation}
The state precesses around $\hat{\mathbf d}^{\,f}(k)$, and its component along this direction is conserved \cite{PhysRevLett.121.130603}. The transverse component determines the largest transition probability accessible to the sector, $A(k)=1-m_k^2$. The initial level spacing is $2\epsilon_i(k)$, so the energy deposited in the sector is
\begin{equation}
\Delta E_k(t)=2\epsilon_i(k)\bigl[1-|\mathcal G_k(t)|^2\bigr].
\label{eq:EkG}
\end{equation}
The corresponding charging fraction, defined as the population of the upper initial level, is $\eta_k(t)=A(k)\sin^2[\epsilon_f(k)t]$. The transition weight $A(k)$ fixes the accessible upper-level population, the final band energy $\epsilon_f(k)$ sets the precession frequency, and the initial band energy $\epsilon_i(k)$ converts that population into stored energy. Appendix~\ref{app:two_level} gives the two-level derivation together with the energy variance and the signal-to-noise ratio of the stored energy.

A real-time zero of Eq.~\eqref{eq:Gk} occurs when $m_{k^*}=0$ and $\epsilon_f(k^*)\neq0$ \cite{Heyl2013}. We denote by $k^*$ the momentum at which the two Hamiltonian directions are orthogonal and by $t_c=\pi/[2\epsilon_f(k^*)]$ the first time at which the return amplitude vanishes. At this time $\eta_{k^*}=1$, so the critical sector occupies the upper initial level with unit probability and its energy reaches a smooth maximum, where the instantaneous sector power $P_k=\partial_t\Delta E_k$ vanishes. The condition $m_{k^*}=0$ expresses orthogonality between the initial and final Hamiltonian directions, while evolution to $t_c$ produces orthogonality between the initial and the evolved state.

For independent sectors the many-body return probability factorizes. The thermodynamic return rate is therefore \cite{Heyl2013}
\begin{equation}
\lambda(t)=-\lim_{N\to\infty}\frac{1}{N}\sum_k\ln|\mathcal G_k(t)|^2.
\label{eq:return_rate}
\end{equation}
Near a simple transverse zero $(k^*,t_c)$, the return probability is quadratic in the momentum and time displacements. Integrating its logarithm over momentum gives a contribution proportional to $|t-t_c|$, which produces a cusp in $\lambda(t)$ \cite{Heyl2013,Heyl2018}. The local expansion is derived in Appendix~\ref{app:dqpt_cusp}.

The energy in Eq.~\eqref{eq:EkG} contains the same transition probability with a different weight. At the critical point the contribution of the critical sector approaches the finite value $2\epsilon_i(k^*)$. For smooth bounded bands the energy and its time derivatives remain integrable over momentum, so the total energy has an ordinary expansion around $t_c$. Complete charging of one sector is therefore compatible with a smooth many-body energy response. The logarithm in the return rate and the bounded kernel in the energy are what make the two observables respond so differently to the same event.

We next consider the long-time-averaged energy, which describes the energy that survives once the relative precession phases have averaged out. The many-body stored energy is the sum of the sector energies of Eq.~\eqref{eq:EkG} over the independent pair sectors, and the stored-energy density follows from dividing that sum by the number of sites. In the thermodynamic limit the momentum sum becomes an integral over positive momenta, and we drop the size subscript, writing $\Delta e(t)=\lim_{N\to\infty}\Delta e_N(t)$. With one independent pair sector for each positive momentum, the average over a long time interval $T$ is
\begin{equation}
\begin{aligned}
\Delta e_\infty
&\equiv \lim_{T\to\infty}\frac{1}{T}\int_0^T \Delta e(t)\,dt \\
&= \frac{1}{\pi}\int_0^\pi \epsilon_i(k)A(k)
\left[
\lim_{T\to\infty}\frac{1}{T}
\int_0^T \sin^2\!\bigl(\epsilon_f(k)t\bigr)\,dt
\right] dk \\
&= \frac{1}{2\pi}\int_0^\pi \epsilon_i(k)A(k)\,dk .
\end{aligned}
\label{eq:plateau}
\end{equation}
For every nonzero mode frequency the oscillatory factor averages to $1/2$, and isolated zero-frequency points have zero measure in the momentum integral. The long-time-averaged energy consequently retains the initial energy weights and the transition amplitudes while the relative precession phases drop out.

For smooth initial data and a gapped final band, the kernel of Eq.~\eqref{eq:plateau} varies smoothly with the quench parameter. Near a gap closing, however, the direction $\hat{\mathbf d}^{\,f}(k)$ can vary rapidly within a shrinking momentum interval, and a nonuniform change of $A(k)$ inside that interval produces different one-sided slopes of the integrated energy. This is the spectral mechanism by which a cusp appears in the long-time-averaged charging response.

The magnitude of that response can be analyzed by splitting the integrand,
\begin{equation}
\begin{aligned}
\Delta e_\infty
&=\frac{1}{2\pi}\int_0^\pi \epsilon_i(k)A(k)\,dk \\
&=\frac{1}{2\pi}\int_0^\pi \epsilon_i(k)\,dk
-\frac{1}{2\pi}\int_0^\pi
\epsilon_i(k)\bigl[1-A(k)\bigr] \,dk .
\end{aligned}
\label{eq:plateau_decomp}
\end{equation}
Since $0\leq A(k)\leq1$ and $\epsilon_i(k)\geq0$, the long-time-averaged energy satisfies
\begin{equation}
\Delta e_\infty
\leq
\frac{1}{2\pi}\int_0^\pi \epsilon_i(k)\,dk
\equiv \Delta e_\infty^{\rm max}.
\label{eq:plateau_bound}
\end{equation}
The bound is reached when
\begin{equation}
\epsilon_i(k)\bigl[1-A(k)\bigr]=0
\end{equation}
for almost every momentum, so every contributing mode must have unit transition weight throughout the integration interval. A DQPT, by comparison, requires $A(k^*)=1$ at a single critical momentum.

At finite time the stored-energy density is
\begin{equation}
\Delta e(t)
=
\frac{1}{\pi}\int_0^\pi
\epsilon_i(k)A(k)
\sin^2[\epsilon_f(k)t] \,dk ,
\label{eq:instant_energy_bound}
\end{equation}
and it obeys
\begin{equation}
\Delta e(t)
\leq
\frac{1}{\pi}\int_0^\pi \epsilon_i(k)\,dk .
\label{eq:instant_global_bound}
\end{equation}
Saturation of this second bound requires unit transition weight and $\sin^2[\epsilon_f(k)t]=1$ at the same time for almost every contributing mode. A flat final band aligns the frequencies and offers a direct way to synchronize the charging, as discussed in Sec.~\ref{subsec:flat_final}.

\section{Results}
\label{sec:results}

We compare spectral and dynamical criticality through the return rate, the stored energy, and the charging power, organizing the comparison around the three control parameters identified above. The anisotropic XY chain is a two-band free-fermion realization of Sec.~\ref{sec:two_band} with analytical transition weights and mode frequencies, and we ask where its stored energy comes from and what fixes the optimum over the final Hamiltonian. A flat final band converts the accessible transition weight into energy delivered at the critical time, and the interacting Ising chain shows how the extraction time separates from the return peak once probability is redistributed among excited states. The weak-quench response is related to the equilibrium static susceptibility, which at fixed system size connects the choice of preparation to an equilibrium measurement. All energies are expressed in units of $J$.

\subsection{Spectral and dynamical criticality in the XY chain}
\label{sec:xy}

As a concrete realization we consider the anisotropic XY chain,
\begin{equation}
H(g,\gamma)=-J\sum_j\left[\frac{1+\gamma}{2}\sigma_j^x\sigma_{j+1}^x
+\frac{1-\gamma}{2}\sigma_j^y\sigma_{j+1}^y-g\sigma_j^z\right].
\label{eq:XYH}
\end{equation}
The parameter $g$ is the transverse field, $\gamma$ is the anisotropy, and $J$ sets the energy scale, with $\sigma_j^{x,y,z}$ the Pauli operators acting on site $j$. We set $J=1$ in the calculations. The Jordan--Wigner transformation maps the spin chain to fermions \cite{Jordan1928}, and Fourier and Bogoliubov transformations then decompose the XY Hamiltonian into independent positive-momentum $(k,-k)$ sectors of fixed fermion parity \cite{PhysRevA.2.1075}.

For an initial or final Hamiltonian, labeled by $\alpha=i,f$, the sector Bloch vector is $\mathbf d^\alpha(k)=(0,2\gamma_\alpha\sin k,2[g_\alpha-\cos k])$. Its positive energy is $\epsilon_\alpha(k)=2M_\alpha(k)$, where
\begin{equation}
M_\alpha(k)=\sqrt{(g_\alpha-\cos k)^2+\gamma_\alpha^2\sin^2k}
\end{equation}
is the half band energy of Hamiltonian $\alpha$, and $g_\alpha$ and $\gamma_\alpha$ are its field and anisotropy. The overlap between the two Hamiltonian directions is
\begin{equation}
m_k=\frac{(g_i-\cos k)(g_f-\cos k)+\gamma_i\gamma_f\sin^2k}
{M_i(k)M_f(k)}.
\label{eq:xy_overlap}
\end{equation}
The numerator of Eq.~\eqref{eq:xy_overlap} determines whether an orthogonal sector is present, while $M_f(k)$ determines the oscillation frequency of that sector.

\begin{figure}[tb]
\centering
\includegraphics[width=\columnwidth]{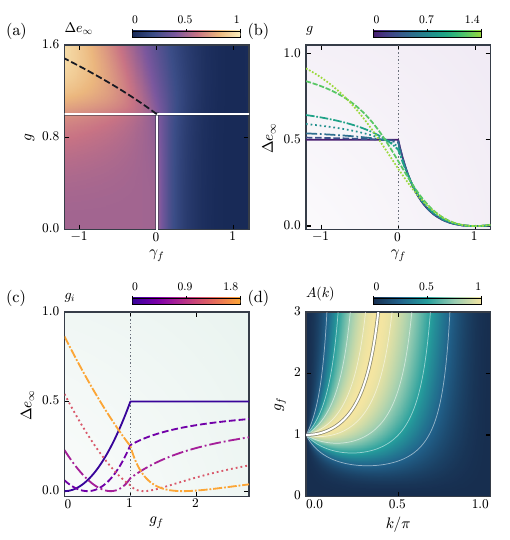}
\caption{Spectral criticality and transition weights in the XY battery. (a) Long-time-averaged energy $\Delta e_\infty$ as a function of the final anisotropy $\gamma_f$ and the field $g$, with $\gamma_i=1$. (b) Cuts of $\Delta e_\infty$ against $\gamma_f$ for several fields $g$. (c) Long-time-averaged energy against the final field $g_f$ on the Ising line $\gamma_i=\gamma_f=1$ for several preparation fields $g_i$. (d) Momentum-resolved transition weight $A(k)$ for $g_i=0$, shown against the final field $g_f$. White and dashed guides mark the spectral and dynamical critical loci discussed in the text.}
\label{fig:xy_global}
\end{figure}

We first fix the field at $g_i=g_f=g$ and change the anisotropy from $\gamma_i=1$ to $\gamma_f$. Figure~\ref{fig:xy_global}(a) maps the resulting long-time-averaged energy over the $(\gamma_f,g)$ plane. The most conspicuous feature is the sharp change along $\gamma_f=0$ for $0<g<1$, which lies exactly on the gapless line of the final XY spectrum \cite{PhysRevA.2.1075}. Approaching this line from either side causes a narrow set of modes around the closing momentum $k_0=\arccos g$ to change their transition weight rapidly, so that the momentum integral of $\epsilon_i(k)A(k)$ acquires a term proportional to $|\gamma_f|$. This is the origin of the nonanalytic slope of the stored energy. The dashed line $\gamma_f=1-g^2$, drawn for $g>1$, has a different meaning. It marks the appearance of an orthogonal momentum sector, and hence the boundary of the DQPT region, even though the final spectrum remains gapped there. The absence of a corresponding ridge or cusp in $\Delta e_\infty$ along this dashed boundary is already a direct visual indication that a DQPT condition and a spectral singularity of the charging energy are different objects.

The distinction can be traced to the way the two observables sample momentum space. A DQPT takes an orthogonal sector at an isolated momentum, which is sufficient to generate a logarithmic singularity in the return-rate integral \cite{Heyl2013}. The long-time-averaged energy contains no logarithm, and it weights $A(k)$ by the smooth factor $\epsilon_i(k)$, so an isolated orthogonal mode has zero measure in that integral. A nonanalytic energy response therefore requires a finite momentum neighborhood to reorganize nonuniformly, as happens when the final gap closes.

The one-dimensional cuts drawn in Fig.~\ref{fig:xy_global}(b) make the change of regularity explicit. For fields $g<1$ the curves have unequal left and right derivatives at $\gamma_f=0$. The discontinuity is strongest away from $g=1$ and weakens as the closing momentum approaches the Brillouin-zone edge. Once $g>1$, changing the sign of $\gamma_f$ no longer closes the final gap, the singular momentum window is absent, and the curves become smooth. They identify the gap closing, rather than the existence of a critical Loschmidt sector, as the mechanism responsible for the cusp in the long-time charging energy.

The singular coefficient is proportional to $\sqrt{1-g^2}$, so the left--right slope jump shrinks continuously as the closing momentum approaches the Brillouin-zone edge. The cusp is the part of the spectral singularity that the quench parameters can act on. Near the critical line the stored energy becomes a sensitive and directed function of the quench parameters, because the sign of $\gamma_f$ selects which of the two limiting slopes applies and the coefficient of that sensitivity is known in closed form.

Spectral criticality therefore enters the optimization as a source of tunability. It marks a region in which the stored energy can be moved by small parameter changes, and it does not by itself imply that the energy is globally maximal there. The size of the maximum is decided by the momentum distribution of the transition weight, while the moment of extraction is decided by the spread of the precession frequencies.

For $|g|<1$, an expansion around the closing momentum gives
\begin{equation}
\frac{d\Delta e_\infty}{d\gamma_f}
=B(g)-\sqrt{1-g^2}\,\operatorname{sgn}(\gamma_f)+o(1).
\label{eq:xy_cusp}
\end{equation}
Here $B(g)$ is the regular contribution to the slope and $\operatorname{sgn}$ is the sign function, while the full expansion is derived in Appendix~\ref{app:xy_cusp}. The singular term comes from the rapid variation of $A(k)$ within a momentum interval whose width is proportional to $|\gamma_f|$, and integrating over that interval produces a contribution linear in $|\gamma_f|$. The initial anisotropy, held at $\gamma_i=1$, contributes to the regular part $B(g)$ of the slope, while the left--right asymmetry comes from the closing mode.

Figure~\ref{fig:xy_global}(c) shows how strongly the stored energy depends on the state from which the quench starts, even when the final Hamiltonian crosses the same equilibrium gap closing at $g_f=1$. We therefore fix $\gamma_i=\gamma_f=1$ and scan the final field for several preparation fields $g_i$. Every curve vanishes at its unquenched point $g_f=g_i$, because no population transfer occurs when the two Hamiltonians coincide. Away from that point the curves differ strongly, since the preparation changes both the transition profile $A(k)$ and the initial excitation energy $\epsilon_i(k)$ that weights each transferred mode. The equilibrium critical point $g_f=1$ is therefore not by itself a universal optimum of the battery. For the special flat initial band at $g_i=0$ the mode weighting simplifies and the long-time-averaged energy becomes
\begin{equation}
\Delta e_\infty(g_f)=
\begin{cases}
g_f^2/2, & 0\leq g_f\leq1,\\
1/2, & g_f\geq1.
\end{cases}
\label{eq:flat_plateau}
\end{equation}
For this preparation the gap-closing point $g_f=1$ marks the onset of the exact plateau $\Delta e_\infty=1/2$, and that value persists even for final Hamiltonians deep in the gapped $g_f>1$ phase. The plateau therefore extends into a region where the final Hamiltonian is fully gapped, and it does not signal that the system stays critical beyond $g_f=1$. The plateau is flat over a wide field interval, so the energy stored there does not respond to small changes of the final field. The value $1/2$ should not be read as the largest accessible one, since it is only half of the bound $\Delta e_\infty^{\rm max}$ of Eq.~\eqref{eq:plateau_bound}, which equals one for a flat initial band. The plateau comes about through a compensation inside the momentum integral, in which the distribution of transition probabilities changes with $g_f$ so as to leave the weighted integral fixed. The other preparation fields make this point especially clear, since their maxima, slopes, and values at $g_f=1$ differ from one another, and in some cases the energy continues to increase after the equilibrium critical field has been crossed. The charging response is thus a property of the quench pair $(H_i,H_f)$ rather than of the final spectrum alone, and the same final critical Hamiltonian can deliver very different energies to different initial states. A peak at or near $g_f=1$ therefore cannot be attributed to the equilibrium critical point without further evidence.

The momentum-resolved transition weight in Fig.~\ref{fig:xy_global}(d) explains the plateau at the microscopic level. For $g_i=0$ the condition $A(k)=1$ selects only a single fully transferable momentum $k^*$, given by $\cos k^*=1/g_f$ when $g_f>1$, and the heavy white curve follows that sector across the plane. The surrounding color field shows that most momenta remain only partially transferred, so the constant energy in Fig.~\ref{fig:xy_global}(c) cannot be read as complete charging of every independent sector. It is an integrated effect of a nonuniform $A(k)$. An analytically solvable case makes the contrast explicit. For $(g_i,g_f)=(-1,1)$ one has $m_k\equiv0$, hence $A(k)=1$ throughout the interior of the Brillouin zone, and the long-time-averaged bound is saturated mode by mode. This preparation is not among the fields scanned in Fig.~\ref{fig:xy_global}(c). Figure~\ref{fig:xy_global}(c) and Fig.~\ref{fig:xy_global}(d) therefore separate a large integrated energy from uniform microscopic transfer. The same measure argument explains why the isolated $A(k^*)=1$ sector that drives a DQPT does not saturate the battery-energy bound. Saturating the integral requires near-unit transition weight over a set of momenta with finite measure, whereas a return-rate zero needs only one orthogonal sector in the thermodynamic continuum.

\begin{figure}[tb]
\centering
\includegraphics[width=\columnwidth]{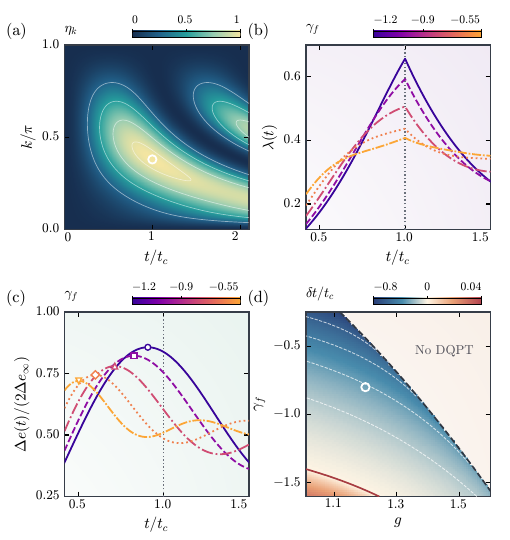}
\caption{Dynamical criticality and charging-time separation for $\gamma_i=1$. (a) Momentum--time map of the charging fraction $\eta_k(t)$ for $\gamma_f=-0.8$. (b) Return rate $\lambda(t)$ against $t/t_c$ for several final anisotropies $\gamma_f$. (c) Stored energy normalized by $2\Delta e_\infty$ for the same quenches. (d) Signed offset $\delta t/t_c=(t_{\max}-t_c)/t_c$ over field and anisotropy, where $t_{\max}$ is the time of the first energy maximum. The red contour marks $t_{\max}=t_c$, and the dashed line separates the DQPT region from the quenches that produce no return zero.}
\label{fig:dqpt_mode}
\end{figure}

We next compare these spectral signatures with gapped anisotropy quenches at $g_i=g_f=1.2$, for which DQPTs occur without an endpoint gap closing \cite{PhysRevB.89.161105,srx7-cpl4}. Figure~\ref{fig:dqpt_mode}(a) resolves the charging fraction $\eta_k(t)$ in momentum and time for the quench $\gamma_i=1\to\gamma_f=-0.8$. The circle marks the unique combination $(k^*,t_c)$ at which the initial and final Bloch directions are orthogonal and the critical sector is transferred completely, so that $\eta_{k^*}=1$. The bright ridge bends across momentum because $\epsilon_f(k)$ is dispersive, so neighboring sectors precess at different frequencies and generally have $A(k)<1$. They therefore reach different maxima at different times and with different amplitudes. A many-body return zero is thus generated by one critical sector without forcing the energy carried by all other sectors to peak simultaneously. The two ingredients remain visible separately, since the transition weight $A(k)$ fixes the maximum transferable fraction of each sector while $\epsilon_f(k)$ fixes its oscillation period. The DQPT condition selects a sector with $A(k^*)=1$ and imposes no constraint on the frequencies or amplitudes of the remaining momenta. The global charging optimum is consequently a collective phase-alignment problem rather than a single-mode orthogonality condition.

Figure~\ref{fig:dqpt_mode}(b) compares the return rates for five final anisotropies after rescaling each time axis by its own first critical time. This rescaling collapses the locations of the cusps to $t/t_c=1$, as expected from the definition of $t_c$, while the cusp heights and the smooth background remain quench dependent. Those differences reflect how many noncritical sectors have appreciable $A(k)$, and how their phases are distributed at the moment when the critical sector becomes orthogonal. A DQPT therefore fixes a singular time through a local condition in momentum space, while the magnitude of the return-rate response is an integrated property of the entire band. This is why normalizing the horizontal axis by $t_c$ aligns the singular locations but cannot collapse the curves themselves. The background and the cusp amplitude retain information about the distribution of transition weights away from $k^*$, so $t_c$ locates the return-rate singularity but does not by itself locate the energy optimum.

Figure~\ref{fig:dqpt_mode}(c) plots the corresponding stored energies after division by the accessible transition-weight integral $2\Delta e_\infty=(1/\pi)\int_0^\pi\epsilon_i(k)A(k)\,dk$. This normalization removes the trivial change in the total accessible transition weight and leaves the degree of dynamical phase alignment visible. In contrast to Fig.~\ref{fig:dqpt_mode}(b), every curve remains smooth through $t/t_c=1$, and the marked first maxima move substantially as $\gamma_f$ is varied. The reason is that the energy is a weighted sum of $\sin^2[\epsilon_f(k)t]$ over all momenta, so maximizing one critical factor at $k^*$ does not maximize that sum unless the relevant frequencies are synchronized. Each sector contributes an analytic trigonometric function, and the momentum integral remains regular at the isolated critical mode. The nonanalyticity of the return rate instead comes from applying the logarithm to a factor that vanishes at $k^*$, which changes the local integrability structure near the zero.

Figure~\ref{fig:dqpt_mode}(d) shows that the resulting separation is systematic rather than peculiar to the representative quench. Throughout most of the gapped DQPT region the offset $\delta t/t_c$ is negative, so the first energy maximum precedes the Loschmidt critical time, while a much narrower region gives a positive offset. The red zero contour identifies the parameters for which the two times coincide. That contour lies inside the DQPT region rather than coinciding with its boundary, so coincidence of the two times is a tuning condition and not a defining consequence of dynamical criticality.

\subsection{Flat-band synchronization}
\label{subsec:flat_final}

We now examine the conditions under which a DQPT time coincides with an energy maximum. A flat initial band makes the factor $\epsilon_i(k)$ uniform in the energy integral, whereas a flat final band makes the precession frequency uniform and aligns the maxima of the contributing sectors.

For an exactly flat final spectrum, $\epsilon_f(k)=\epsilon_0$ with constant $\epsilon_0$, the energy becomes
\begin{equation}
\begin{aligned}
\Delta e(t)
&=
\Delta e_{\rm sync}\sin^2(\epsilon_0 t),\\
\Delta e_{\rm sync}
&=
\frac{1}{\pi}\int_0^\pi \epsilon_i(k)A(k)\,dk ,
\end{aligned}
\label{eq:flat_final_energy}
\end{equation}
where $\Delta e_{\rm sync}$ is the largest energy that this quench can store. Comparison with Eq.~\eqref{eq:plateau} shows that the synchronized value is exactly twice the long-time-averaged energy, so flattening the final band recovers the whole oscillatory budget that phase averaging discards. The return rate is
\begin{equation}
\lambda(t)
=
-\frac{1}{2\pi}\int_0^\pi
\ln\!\left[1-A(k)\sin^2(\epsilon_0 t)\right]dk .
\label{eq:flat_final_rate}
\end{equation}
When a critical momentum satisfies $A(k^*)=1$, the dynamical transition times are $t_c^{(n)}=(2n+1)\pi/(2\epsilon_0)$ for nonnegative integer $n$. At each of these times the energy reaches $\Delta e_{\rm sync}$, its global maximum for the specified quench.

The energy remains smooth at the synchronized maximum. Expanding Eq.~\eqref{eq:flat_final_energy} gives $\Delta e(t)=\Delta e_{\rm sync}[1-\epsilon_0^2(t-t_c)^2+O((t-t_c)^4)]$, while the return rate retains its cusp. Flatness aligns the phases of the different sectors, and the value of $\Delta e_{\rm sync}$ still depends on $A(k)$, with complete charging of the battery obtained only when the contributing modes also carry unit transition weight.

To determine the effect of weak dispersion we write $\epsilon_f(k)=\epsilon_*[1+\delta_k]$, where $\epsilon_*=\epsilon_f(k^*)$ is the band energy at the critical momentum and $\delta_k$ is the relative deviation from it, assumed to satisfy $|\delta_k|\ll1$ over the momenta that carry appreciable energy weight. At $t_c=\pi/(2\epsilon_*)$ the normalized energy is
\begin{equation}
\frac{\Delta e(t_c)}{\Delta e_{\rm sync}}
=
1-\frac{\pi^2}{4}
\left\langle \delta_k^2\right\rangle_W
+O(\delta_k^4).
\label{eq:near_flat_sync}
\end{equation}
Here $W(k)=\epsilon_i(k)A(k)$ is the energy weight of momentum $k$, and the weighted average of a momentum-dependent quantity $X(k)$ is defined as
\begin{equation*}
\langle X\rangle_W=
\frac{\int_0^\pi W(k)X(k)\,dk}{\int_0^\pi W(k)\,dk}.
\end{equation*}
The energy loss at $t_c$ is second order in the relative frequency spread. Expanding the stationarity condition $\partial_t\Delta e(t_{\max})=0$ gives the corresponding timing shift,
\begin{equation}
\frac{t_{\max}}{t_c}=1-\langle\delta_k\rangle_W+O(\delta_k^2).
\label{eq:near_flat_time}
\end{equation}
An asymmetric weighted dispersion therefore shifts the maximum at first order, even when the reduction of the energy at $t_c$ remains quadratic.

\begin{figure}[tb]
\centering
\includegraphics[width=\columnwidth]{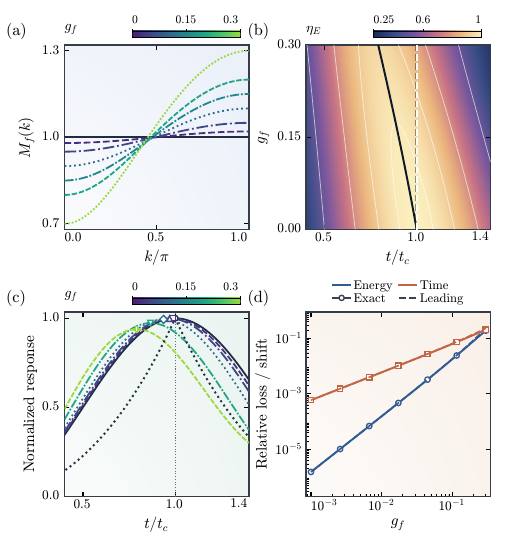}
\caption{Flat-band synchronization on the Ising line for $g_i=1.3$. (a) Final half band energy $M_f(k)$ against momentum for several final fields $g_f$. (b) Normalized stored energy $\eta_E(t)=\Delta e(t)/(2\Delta e_\infty)$ over time and final field, with the ridge of first maxima $t_{\max}$ in solid black and the critical time $t_c$ marked by the dashed vertical line. (c) Time cuts of the normalized response together with the return rate of the exactly flat quench. (d) Energy loss $1-\eta_E(t_c)$ and timing shift $1-t_{\max}/t_c$ as the final band approaches flatness, compared with the leading predictions of Eqs.~\eqref{eq:near_flat_sync} and \eqref{eq:near_flat_time}.}
\label{fig:flat_final_sync}
\end{figure}

Figure~\ref{fig:flat_final_sync}(a) displays the quantity that this section varies deliberately. At $g_f=0$ the half band energy satisfies $M_f(k)=1$ for every momentum, so all independent sectors accumulate the same dynamical phase. Increasing $g_f$ tilts and curves the band, widening the spread of $M_f(k)$ and hence of the precession frequencies $\epsilon_f(k)=2M_f(k)$. The dispersion curves turn the qualitative notion of band flatness into a specific dynamical quantity, the spread of precession frequencies. What matters for synchronization is then a small frequency spread over the momenta that carry appreciable weight $W(k)=\epsilon_i(k)A(k)$ rather than a small bandwidth in an absolute sense. Dispersion in weakly populated sectors has little effect on the total energy, whereas comparable dispersion in strongly weighted sectors spreads the relative phases at once.

Figure~\ref{fig:flat_final_sync}(b) shows the consequence for the battery. At $g_f=0$ the black ridge of $t_{\max}$ intersects the vertical critical-time line at $t/t_c=1$, and the normalized energy reaches unity there. Throughout this subsection we normalize the stored energy by its long-time value and write $\eta_E(t)=\Delta e(t)/(2\Delta e_\infty)$ for the resulting normalized response, which approaches unity only when the weighted sectors reach their maxima simultaneously and tends to $1/2$ once their precession phases spread apart. As dispersion grows, the ridge moves systematically away from the DQPT line and the high-energy contours become less concentrated around $t/t_c=1$. Different momentum sectors now arrive at their maxima at different times, so their contributions cannot add coherently at the critical-sector time. The reduction is therefore an averaging effect within the closed unitary dynamics, and no energy is dissipated to an environment. Since this loss is generated by the dispersion of the final band, it is controlled by the post-quench Hamiltonian.

The time cuts in Fig.~\ref{fig:flat_final_sync}(c) show the same physics in one dimension. The exactly flat quench has a smooth energy maximum at the same time as the sharp return-rate cusp, which emphasizes that temporal coincidence does not make the two observables locally equivalent. With increasing $g_f$ the energy maximum moves and decreases, while the DQPT time stays fixed by the orthogonal sector after the chosen normalization. A flat final band therefore lets the DQPT time coincide with the energy maximum, which is the sense in which a critical time can be made useful for charging. Even at exact synchronization the return rate and the energy keep different local shapes, since the former is nonanalytic and the latter has an ordinary quadratic maximum.

Figure~\ref{fig:flat_final_sync}(d) quantifies how rapidly this synchronization is lost. On logarithmic axes the energy deficit $1-\eta_E(t_c)$ follows the quadratic prediction of Eq.~\eqref{eq:near_flat_sync}, whereas the timing error $1-t_{\max}/t_c$ follows the leading linear dependence of Eq.~\eqref{eq:near_flat_time}. A weak residual bandwidth can shift the optimal charging time appreciably before it causes a comparably large loss of the energy stored at the nominal DQPT time, so timing is the more sensitive diagnostic of imperfect flatness. For a protocol that extracts at a fixed time, the bandwidth requirement is therefore set by the timing error rather than by the energy loss, since a small residual bandwidth leaves most of the synchronized energy intact while moving the optimal extraction time away from $t_c$.

\subsection{Energy redistribution in the interacting chain}
\label{sec:interacting}

To go beyond the independent-sector framework of Sec.~\ref{sec:two_band} we consider an interacting transverse-field Ising chain with next-nearest-neighbor coupling,
\begin{equation}
\begin{split}
H(g,\kappa)={}&-J\sum_{j=1}^N\sigma_j^z\sigma_{j+1}^z
-\kappa J\sum_{j=1}^N\sigma_j^z\sigma_{j+2}^z\\
&-gJ\sum_{j=1}^N\sigma_j^x.
\end{split}
\label{eq:nnn_model}
\end{equation}
Here $g$ is again the transverse field, $J$ sets the energy scale, and $\kappa$ is the dimensionless strength of the next-nearest-neighbor coupling. We impose periodic boundary conditions and use a rotated spin convention relative to Eq.~\eqref{eq:XYH}. Positive $\kappa$ gives the next-nearest-neighbor term the same ferromagnetic sign as the nearest-neighbor coupling, so it reinforces the Ising order. This differs from the standard frustrated ANNNI model, in which the nearest- and next-nearest-neighbor interactions compete \cite{Selke}. Throughout the interacting calculations we set $\kappa=0.2$. At nonzero transverse field this model has no free-fermion factorization. Because the interactions remain finite ranged, the model lies within the setting for which Lieb and Robinson established a finite bound on the propagation velocity of information \cite{LiebRobinson}. The return probability and the energy nevertheless probe different properties of the evolving many-body state.

We perform exact diagonalization for $N=8,10,12,14$ in the zero-momentum, even spin-flip sector that contains the initial ground state. Both Hamiltonians preserve this sector, and all eigenstates in the corresponding block are retained. The finite-size dynamics, the charging power, and the long-time-averaged energy are calculated from the same spectral decomposition. Details of the symmetry reduction and of the time averaging are given in Appendix~\ref{app:ed}.

\begin{figure}[tb]
\centering
\includegraphics[width=\columnwidth]{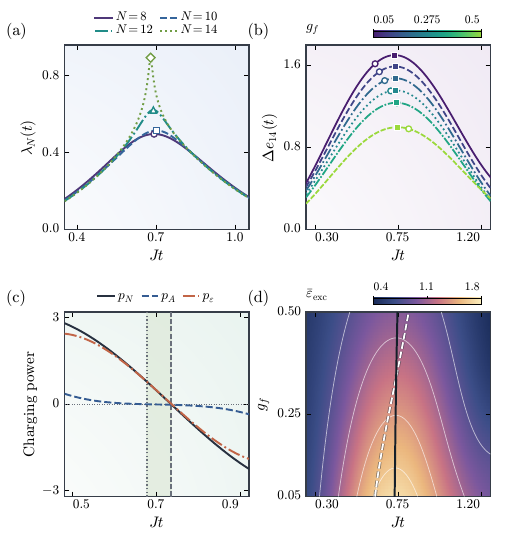}
\caption{Interacting battery dynamics for $g_i=1.8$ and $\kappa=0.2$. (a) Finite-size return rate $\lambda_N(t)$ for $N=8,10,12,14$ at $g_f=0.2$. (b) Stored energy $\Delta e_{14}(t)$ for several final fields at $N=14$, with open circles at the return-rate peaks and filled squares at the energy maxima. (c) Charging power per site $p_N$ at $g_f=0.2$, resolved into the population term $p_A$ and the redistribution term $p_\varepsilon$. (d) Conditional excitation energy $\bar\varepsilon_{\mathrm{exc},14}$ over time and final field, with the solid black curve marking the energy maximum and the dashed white curve the return-rate peak.}
\label{fig:nnn_dynamics}
\end{figure}

For the representative quench $g_i=1.8\to g_f=0.2$ at $\kappa=0.2$, Fig.~\ref{fig:nnn_dynamics}(a) shows the finite-size return rate $\lambda_N(t)$ for $N=8,10,12,14$. The peak becomes sharper and higher as $N$ increases, which is the expected finite-size signature of an increasingly strong suppression of the Loschmidt overlap \cite{Heyl2013}. Its time, however, is not monotonic in $N$, so we use the peak as a finite-size dynamical marker rather than extrapolating its position from a single size. For the battery problem this curve provides the time at which the depletion of the initial state is strongest, and it does not yet indicate where in the excited spectrum the transferred population resides. The peak height rises with $N$ over the sizes available here, which points to an increasingly strong suppression of the overlap with the initial ground state near this time, but that information is binary with respect to the excited manifold.

Figure~\ref{fig:nnn_dynamics}(b) supplies the missing information by plotting the stored energy at fixed $N=14$ for six final fields. Open circles mark the return-rate peaks and filled squares mark the actual energy maxima, and their relative positions change across the scan. For the representative $g_f=0.2$ quench the open circle lies on the rising side of the energy curve, whereas at other fields the two markers approach or pass one another. The height of the energy maximum also changes strongly with $g_f$. Even after momentum-sector factorization is lost, the return peak and the charging optimum therefore need not coincide. The reason is especially transparent in an interacting many-level system. A small overlap with the initial ground state fixes the total excited-state population, while the stored energy depends in addition on how that population is distributed over the excitation energies. Two states can thus have almost the same Loschmidt overlap and still store different energies if their excited-state weights are centered at different parts of the initial spectrum.

The charging power per site \cite{Binder2015,Campaioli2017} resolves this freedom into its parts. We evaluate it as
\begin{equation}
p_N(t)=\frac{i}{N}\langle\psi(t)|[H_f,H_i]|\psi(t)\rangle,
\label{eq:interacting_power}
\end{equation}
and Fig.~\ref{fig:nnn_dynamics}(c) separates it into two physically different processes. The dashed population contribution $p_A$ measures the rate at which probability leaves or returns to the initial ground state, whereas $p_\varepsilon$ measures the motion of already excited population toward higher or lower initial-energy levels. At the return-rate maximum $A_N$ is stationary and hence $p_A=0$, yet the total power remains positive because $p_\varepsilon>0$. The shaded interval between the return peak and the energy maximum is therefore a period of continued charging without further net depletion of the initial ground state. Charging stops only when the redistribution contribution can no longer compensate the population term and the total power crosses zero. For this quench the energy maximum therefore falls after the return-rate peak, and the length of the interval between the two is set by how far the excited population can be moved upward in the initial spectrum. The ordering is not fixed, however, and Fig.~\ref{fig:nnn_dynamics}(d) shows the two curves crossing as the final field is varied. This decomposition gives a dynamical meaning to the separation of the markers in Fig.~\ref{fig:nnn_dynamics}(b). The return peak signals the end of net ground-state depletion, whereas the energy maximum signals the later cancellation of all channels that change the mean initial energy, so the interval between them is a genuine intra-excited-manifold charging stage.

Figure~\ref{fig:nnn_dynamics}(d) shows that the same mechanism governs the whole final-field range rather than being tied to the single cut $g_f=0.2$. Its color scale is the conditional excitation energy $\bar\varepsilon_{\mathrm{exc},14}$, the mean initial-spectrum energy carried by the population that has already left the ground state. Between the dashed white curve and the solid black curve, this conditional energy generally continues to rise after the return peak. The crossing of the two curves as $g_f$ varies shows that the ordering of the two characteristic times is itself tunable. The map therefore provides a many-body counterpart of the free-fermion timing map in Fig.~\ref{fig:dqpt_mode}(d), with one important difference in the underlying cause. In the interacting problem the separation of the two times is generated by redistribution across a hierarchy of many-body excitation energies rather than by unsynchronized independent momenta. The broad region of increasing $\bar\varepsilon_{\mathrm{exc},14}$ beyond the return marker shows that this behavior is not tied to the single field $g_f=0.2$ but holds across most of the scanned range. We expect the same mechanism in interacting batteries whose excited manifold is dense enough for probability to be rearranged while the total excited-state weight stays nearly constant.

To describe the underlying redistribution we expand the state in the eigenbasis of $H_i$ as $|\psi(t)\rangle=\sum_n a_n(t)|n_i\rangle$, where $|n_i\rangle$ are the eigenstates of $H_i$ with energies $E_n^i$ and $a_n(t)$ are the time-dependent amplitudes, and we define the populations $p_n(t)=|a_n(t)|^2$. The nondegenerate initial ground state is labeled by $n=0$. The total probability in the excited-state subspace is
\begin{equation}
A_N(t)=1-|\mathcal G(t)|^2=\sum_{n>0}p_n(t),
\end{equation}
while the stored-energy density is
\begin{equation}
\Delta e_N(t)=\frac{1}{N}\sum_{n>0}
(E_n^i-E_0^i)p_n(t).
\label{eq:app_energy_population}
\end{equation}
For $A_N(t)>0$ the conditional excitation energy per site is
\begin{equation}
\bar\varepsilon_{\mathrm{exc},N}(t)=
\frac{\sum_{n>0}(E_n^i-E_0^i)p_n(t)}
{N A_N(t)}.
\label{eq:app_conditional_energy}
\end{equation}
These quantities provide the decomposition
\begin{equation}
\Delta e_N(t)
=
A_N(t)\,\bar\varepsilon_{\mathrm{exc},N}(t).
\label{eq:activation_energy_decomp}
\end{equation}
Here $A_N$ measures the depletion of the initial ground state and $\bar\varepsilon_{\mathrm{exc},N}$ measures the mean energy of the excited population in the initial spectrum. In a many-level system that mean energy can change through redistribution among the excited states even when the ground state is already strongly depleted.

The excited-state probability and the return rate obey
\begin{equation}
A_N(t)=1-e^{-N\lambda_N(t)},
\end{equation}
so their smooth maxima occur at the same times. At $g_f=0.2$ the conditional excitation energy continues to increase beyond the return-rate peak. Differentiating Eq.~\eqref{eq:activation_energy_decomp} gives
\begin{equation}
p_N(t)=
\dot A_N(t)\bar\varepsilon_{\mathrm{exc},N}(t)
+
A_N(t)\dot{\bar\varepsilon}_{\mathrm{exc},N}(t).
\label{eq:app_power_decomposition}
\end{equation}
The two terms define $p_A=\dot A_N\bar\varepsilon_{\mathrm{exc},N}$ and $p_\varepsilon=A_N\dot{\bar\varepsilon}_{\mathrm{exc},N}$, and at a smooth maximum of $A_N$ the population term vanishes. Figure~\ref{fig:nnn_dynamics}(c) shows that the redistribution term remains positive at this time and supplies the total charging power, which vanishes only later, when the energy turns. At times with a positive limiting return rate, $A_N$ approaches one as the size increases and the conditional energy takes over the description of the distribution within the excited-state subspace.

We also scan the final field at fixed $g_i=1.8$ to compare the long-time-averaged energy with an equilibrium spectral response. The results are shown in Fig.~\ref{fig:int_steady} and discussed in Appendix~\ref{app:ed}. The derivative has a broad size-dependent maximum in the comparison interval, and the equilibrium fidelity susceptibility \cite{YouLiGu2007,Gu2010Fidelity} peaks in a nearby field region with a different size dependence and a different peak position. These finite-quench observables weight the spectrum differently. A direct relation to an equilibrium response is obtained in the weak-quench limit.

\subsection{Weak-quench response and static susceptibility}
\label{sec:weak_quench}

We now consider a small change of the control parameter and evaluate the leading contribution to the long-time-averaged energy at fixed finite size. Let $H(g)$ be a smooth family of Hamiltonians with a nondegenerate ground state, and write $V=\partial_gH$ for the derivative of the Hamiltonian with respect to the control parameter, $V_{n0}=\langle n|V|0\rangle$ for its matrix element between the ground state $|0\rangle$ and an excited eigenstate $|n\rangle$, and $\omega_{n0}=E_n-E_0>0$ for the corresponding excitation energy. For a sudden quench $g\to g+\delta g$ of amplitude $\delta g$, the stored-energy density is
\begin{equation}
\overline{\Delta e}_N
=
(\delta g)^2\chi_V^{(N)}(g)
+
O((\delta g)^3),
\label{eq:small_quench_energy}
\end{equation}
where
\begin{equation}
\chi_V^{(N)}(g)
=
\frac{2}{N}
\sum_{n>0}
\frac{|V_{n0}|^2}{\omega_{n0}} .
\label{eq:static_chi}
\end{equation}
The coefficient $\chi_V^{(N)}$ is the zero-temperature static spectral susceptibility per site associated with the quenched operator. For a Hamiltonian of the form $H=H_0+gV$ it equals $-N^{-1}d^2E_0/dg^2$, and for the transverse-field form $H=H_0-gM$ it reduces to the ordinary static susceptibility $N^{-1}d\langle M\rangle/dg$. The derivation uses many-body eigenstates directly and applies to integrable and interacting systems alike at fixed finite size. Appendix~\ref{app:small_quench} gives the full perturbative derivation and distinguishes this $1/\omega$ spectral response from the $1/\omega^2$ fidelity susceptibility \cite{YouLiGu2007,Gu2010Fidelity}.

\begin{figure}[tb]
\centering
\includegraphics[width=\columnwidth]{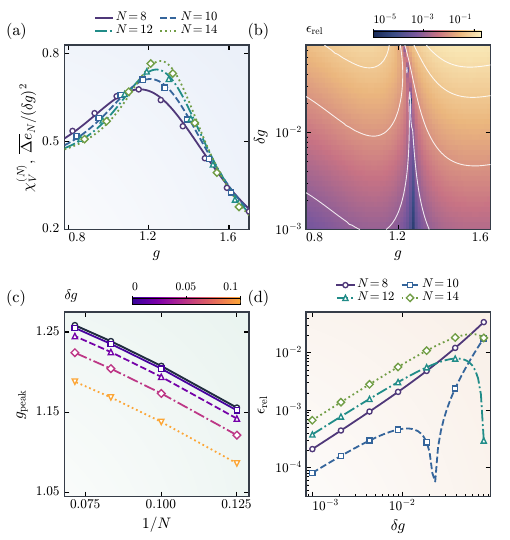}
\caption{Weak-quench response at $\kappa=0.2$. (a) Static spectral susceptibility $\chi_V^{(N)}(g)$ for four sizes, together with the scaled long-time-averaged energy $\overline{\Delta e}_N/(\delta g)^2$ obtained from quenches of amplitude $\delta g=0.02$. (b) Relative mismatch $\epsilon_{\mathrm{rel}}$ over field $g$ and quench amplitude $\delta g$ for $N=14$. (c) Peak field $g_{\mathrm{peak}}$ against inverse size for four quench amplitudes, compared with the peak of $\chi_V^{(N)}$. (d) Mismatch $\epsilon_{\mathrm{rel}}$ against quench amplitude at $g=1.2$ for four system sizes.}
\label{fig:small_quench}
\end{figure}

Figure~\ref{fig:small_quench}(a) gives a direct numerical test of Eq.~\eqref{eq:small_quench_energy}. Continuous curves show the static spectral susceptibility calculated from the equilibrium eigenstates of the initial Hamiltonian, and symbols show the values obtained independently from the long-time-averaged energy after a finite quench of amplitude $\delta g=0.02$. Their close overlap over the full field interval shows that the leading charging response is already encoded in the equilibrium matrix elements $|V_{n0}|^2$ and excitation gaps $\omega_{n0}$. The peak grows and shifts with $N$, and the two calculations track the same finite-size evolution. The remaining separation between symbols and curves is the finite-amplitude correction represented by the $O((\delta g)^3)$ term. The two data sets come from different information. The symbols use the full post-quench diagonal ensemble, whereas the curves require only equilibrium matrix elements of the initial Hamiltonian. Their agreement therefore supports the susceptibility relation as a quantitative predictor of the weak-quench response at fixed $N$ and small $\delta g$.

Figure~\ref{fig:small_quench}(b) resolves the relative mismatch $\epsilon_{\mathrm{rel}}=|\overline{\Delta e}_N/(\delta g)^2\chi_V^{(N)}-1|$ over both initial field and quench amplitude for $N=14$. The dark region at small $\delta g$ shows the uniform approach to the perturbative limit, and the mismatch grows as the quench is made larger. The narrow low-error valley at finite amplitude results from partial cancellation among higher-order corrections and therefore moves with $\delta g$. It is not a new critical line. The map identifies where the simple equilibrium predictor is quantitatively reliable and where a genuinely finite-quench calculation becomes necessary. The error grows most rapidly where the low-energy spectrum is dense, because the expansion parameter is effectively the quench amplitude measured against the relevant finite-size gaps. A fixed value of $\delta g$ therefore need not have the same perturbative quality throughout the field scan.

Figure~\ref{fig:small_quench}(c) focuses on the location of the response maximum rather than its height. For every finite size, reducing $\delta g$ brings the peak field extracted from $\overline{\Delta e}_N/(\delta g)^2$ toward the peak of $\chi_V^{(N)}$. Larger quenches displace the maximum because higher-order terms sample the spectrum with additional matrix-element and energy-denominator combinations. At fixed $N$ and sufficiently small $\delta g$ the equilibrium susceptibility therefore predicts both the magnitude of the weak-quench response and the field at which it is largest. Since $\chi_V^{(N)}$ is built from equilibrium matrix elements and gaps, that field can be found without running the nonequilibrium protocol, and the displacement at larger amplitudes measures how far the working point has moved beyond the linear response regime. Within this regime the choice of preparation is an equilibrium problem, and the quench amplitude only sets how far the quench may be moved away from the equilibrium prediction.

Figure~\ref{fig:small_quench}(d) examines the same convergence at a fixed field $g=1.2$. For each $N$ the relative error tends to zero as $\delta g\to0$, in agreement with Eq.~\eqref{eq:small_quench_energy}. The nonmonotonic dip for some sizes, most visibly $N=10$, reflects the cancellation of subleading terms at a finite amplitude rather than a failure of the perturbative expansion. As $N$ increases and the low-lying gap decreases, the amplitude range over which the leading term dominates can become narrower, which is why the weak-quench limit has to be taken at fixed $N$ before any large-$N$ scaling is discussed. The two limits need not commute near criticality, since a larger $N$ reduces the smallest excitation gaps and can turn a previously small quench into a nonperturbative one. This finite-size dependence is an explicit numerical warning against inferring thermodynamic scaling from a fixed quench amplitude without first verifying convergence in $\delta g$.

The static susceptibility entering Eq.~\eqref{eq:small_quench_energy} weights each excitation by $|V_{n0}|^2/\omega_{n0}$, whereas the fidelity susceptibility contains $|V_{n0}|^2/\omega_{n0}^2$ \cite{YouLiGu2007,Gu2010Fidelity,ZanardiPaunkovic2006}. The extra inverse gap makes the fidelity susceptibility more sensitive to the softest excitations and gives it different units and different finite-size scaling. Appendix~\ref{app:small_quench} derives both expressions from the same perturbative matrix elements and rewrites them as different inverse-frequency moments of a common spectral density. The two quantities are therefore closely related, and the difference in the inverse-gap power makes them unsuitable substitutes for one another.

Because the weak-quench response is proportional to the curvature of the ground-state energy, it also reports the order of the transition that the control parameter drives the system through \cite{Sachdev2011,MatthiasVojta_2003}. That order is fixed by the lowest derivative of the ground-state energy that becomes singular at the transition. A continuous transition makes the second derivative diverge, so the charging peak sharpens and grows with the system size. A first-order transition instead makes the first derivative jump, which appears in the curvature as a spike of finite weight rather than as a power-law divergence. The two orders are therefore separated by the size dependence of the charging peak rather than by its presence. The measurement is dynamical, but the quantity it accesses is the same ground-state curvature on which the equilibrium classification rests. The identification assumes a Hamiltonian linear in the control parameter, so that Eq.~\eqref{eq:s_chi_curvature} holds without the additional term of Eq.~\eqref{eq:s_nonlinear}, and it requires the small-$\delta g$ limit to be taken at each size before the size dependence is read off.

\section{Conclusion}
\label{sec:conclusion}

Spectral and dynamical criticality affect the stored energy in different ways, and neither locates the optimum by itself. Spectral criticality supplies tunability. A closing gap changes the transition weight only within a narrow window in momentum space, and in the XY chain the anisotropy-driven closing leaves a cusp whose coefficient is $\sin k_0=\sqrt{1-g^2}$ at the closing momentum $k_0=\arccos g$. The stored energy then moves sharply and directionally with the quench parameters, so the critical region is easy to steer, but the energy there is not automatically the largest. Dynamical criticality marks a different event. A DQPT flips a single momentum sector completely while the integrated energy stays smooth, because the energy sums bounded transition weights while the return rate takes their logarithm, so a return zero occurs when the evolving state becomes orthogonal to its initial state. Coincidence of the two times requires the final precession frequencies to agree, which here happens for a flat band at $g_f=0$ rather than at the critical field. The interacting chain has no momentum factorization, and the two times there can be separated by a tunable interval.

Optimization therefore reduces to two objects, the momentum distribution of the transition weights and the spread of the final precession frequencies, which the three control parameters act on. The long-time-averaged energy is bounded, the bound is reached only when the transition weight is near unity at almost every momentum, and the deficit is the initial excitation energy weighted by the squared overlap of the two Bloch directions. An exactly flat final band turns the bound into delivered energy and doubles the value at the critical time, and a preparation such as $(g_i,g_f)=(-1,1)$ saturates the bound against a critical final Hamiltonian.

At fixed size and small quench amplitude the weak-quench problem closes onto equilibrium, since the stored energy is then proportional to the static spectral susceptibility of the quenched operator. That susceptibility weights each excitation by one inverse gap, whereas the fidelity susceptibility uses two and therefore overweights the softest mode, and at fixed size it locates the field of largest response. Because the same susceptibility is the curvature of the ground-state energy, its peak behavior also classifies the transition whenever the control parameter enters linearly, diverging for a continuous one and forming a finite-weight spike for a first-order one. The stored energy a quench delivers is thus a quantity the quench parameters control, rather than a fixed consequence of the equilibrium phase diagram.

\begin{acknowledgments}
This work was supported by the National Natural Science Foundation of China under Grant No.~12575009.
\end{acknowledgments}

\section*{Data Availability}
The numerical data and source code supporting this work are available from the authors upon reasonable request.

\appendix
\input{appendix}

\bibliography{references}

\end{document}

%% file: appendix.tex
\section{Two-level dynamics, energy, and fluctuations}
\label{app:two_level}

\subsection{Return amplitude and transition probability}

In this appendix we derive the return amplitude, the stored energy, and the fluctuations of an independent two-level sector. We suppress the momentum label and write the initial and final sector Hamiltonians as $h_i=\mathbf d_i\cdot\bm\sigma$ and $h_f=\mathbf d_f\cdot\bm\sigma$, with scalar offsets removed, where $\mathbf d_{i,f}$ are the corresponding Bloch vectors and $\bm\sigma$ collects the Pauli matrices. Their positive energy scales are $\epsilon_i$ and $\epsilon_f$.

Using $(\hat{\mathbf d}_f\cdot\bm\sigma)^2=\mathbb I$, with $\mathbb I$ the identity matrix, the evolution operator is
\begin{equation}
e^{-ih_ft}=\cos(\epsilon_ft)\mathbb I-i\sin(\epsilon_ft)\hat{\mathbf d}_f\cdot\bm\sigma.
\label{eq:s_exp}
\end{equation}
The lower initial eigenstate $|u_-^i\rangle$ has Bloch vector $-\hat{\mathbf d}_i$, which gives $\langle u_-^i|\hat{\mathbf d}_f\cdot\bm\sigma|u_-^i\rangle=-\hat{\mathbf d}_i\cdot\hat{\mathbf d}_f$. Substitution into Eq.~\eqref{eq:s_exp} gives the amplitude in Eq.~\eqref{eq:Gk}, with $m=\hat{\mathbf d}_i\cdot\hat{\mathbf d}_f$ the overlap of the two unit Bloch vectors.

The two terms of the amplitude are real and imaginary, respectively, and their squared modulus is $|\mathcal G|^2=1-A\sin^2(\epsilon_ft)$, where $A=1-m^2$ is the transition weight introduced in the main text. Since the sector contains two initial energy eigenstates, the upper-level population is $p(t)=1-|\mathcal G(t)|^2$. Multiplying this population by the level spacing $2\epsilon_i$ yields Eq.~\eqref{eq:EkG}. For a many-level battery the analogous excited-state probability is distributed among different energy levels, which leads to the conditional-energy factor of Eq.~\eqref{eq:activation_energy_decomp}.

\subsection{Geometric derivation and the mode power}

The same charging law follows from the precession of the Bloch vector. The initial vector is $\mathbf r_0=-\hat{\mathbf d}_i$, and evolution under $h_f$ rotates it around $\hat{\mathbf d}_f$ through an angle $2\epsilon_ft$. Applying Rodrigues' formula gives
\begin{equation}
\begin{split}
\mathbf r(t)={}&\mathbf r_0\cos(2\epsilon_ft)
+(\hat{\mathbf d}_f\times\mathbf r_0)\sin(2\epsilon_ft)\\
&+\hat{\mathbf d}_f(\hat{\mathbf d}_f\cdot\mathbf r_0)
[1-\cos(2\epsilon_ft)].
\end{split}
\label{eq:s_rotation}
\end{equation}
Taking the scalar product with $\mathbf d_i$ removes the cross-product term, because $\mathbf r_0$ and $\mathbf d_i$ are collinear. The remaining terms give $\langle h_i\rangle_t=-\epsilon_i+2\epsilon_iA\sin^2(\epsilon_ft)$, and subtracting the initial energy $-\epsilon_i$ reproduces Eq.~\eqref{eq:EkG}. In this representation $A$ is the squared component of the initial direction perpendicular to the final rotation axis.

The instantaneous sector power follows from differentiation,
\begin{equation}
P(t)=2\epsilon_i A\epsilon_f\sin(2\epsilon_ft).
\label{eq:s_power}
\end{equation}
For $m=0$, complete transfer to the upper initial level occurs at $\epsilon_ft=(2n+1)\pi/2$. At these times the stored energy is maximal and the power vanishes, and the change in the sign of $P$ describes charging followed by the coherent return of energy. For $m\neq0$ the energy has the same sequence of turning points with a reduced maximum population. A power zero therefore identifies a turning point of the energy, while a return zero additionally requires orthogonal Hamiltonian directions.

\subsection{Energy variance and the signal-to-noise ratio}

A projective measurement of $h_i$ gives $-\epsilon_i$ with probability $1-p$ and $+\epsilon_i$ with probability $p$, where $p$ is the upper-level population. The first and second moments are $-\epsilon_i+2\epsilon_ip$ and $\epsilon_i^2$, so the energy variance is
\begin{equation}
(\delta E)^2=4\epsilon_i^2p(1-p).
\label{eq:s_variance}
\end{equation}
It is largest for equal populations and vanishes at both initial energy eigenstates, which correspond to an empty sector at $p=0$ and a fully charged sector at $p=1$.

For nonzero variance we define the signal-to-noise ratio as $R_{\mathrm{SNR}}=\Delta E/\delta E$. Equation~\eqref{eq:s_variance} gives $R_{\mathrm{SNR}}^2=p/(1-p)$, whose limits are zero as $p\to0$ and infinity as $p\to1$. The ratio also obeys
\begin{equation}
\ln(1+R_{\mathrm{SNR}}^2)=-\ln|\mathcal G|^2.
\label{eq:s_snr_log}
\end{equation}
At full charging the finite stored energy is measured with zero variance, which produces the divergence of $R_{\mathrm{SNR}}$. Equation~\eqref{eq:s_snr_log} applies to each independent sector, and summing its logarithmic form over momenta gives the return rate, whereas an energy signal-to-noise ratio constructed after summing all sectors is a different quantity.

Integrating Eq.~\eqref{eq:s_snr_log} over positive momenta gives
\begin{equation}
\lambda_{\mathrm{snr}}(t)
=\frac{1}{2\pi}\int_0^\pi
\ln[1+R_{\mathrm{SNR},k}^{2}(t)]\,dk
=\lambda(t).
\end{equation}
The logarithmic fluctuation measure therefore retains the Loschmidt cusp even when the mean stored energy varies smoothly.

\section{Local origin of the return-rate cusp}
\label{app:dqpt_cusp}

We derive the singular contribution to the return rate by expanding the Loschmidt amplitude around a simple transverse zero $(k^*,t_c)$. Let $q=k-k^*$ be the momentum displacement and $\tau=t-t_c$ the time displacement. To first order the amplitude is $\mathcal G(q,\tau)=\alpha q+\beta\tau+O(q^2,q\tau,\tau^2)$, where $\alpha$ and $\beta$ are the derivatives of the amplitude with respect to $q$ and $\tau$. At a transverse zero the real and imaginary parts of these two derivative directions are linearly independent.

The squared modulus then contains a mixed term proportional to $q\tau$, and completing the square gives
\begin{equation}
|\mathcal G(q,\tau)|^2=a(q-u\tau)^2+b\tau^2+O(3).
\label{eq:s_localnormal}
\end{equation}
The coefficients are $a=|\alpha|^2$, $u=-\operatorname{Re}(\alpha^*\beta)/a$, and $b=|\beta|^2-[\operatorname{Re}(\alpha^*\beta)]^2/a$. Both $a$ and $b$ are positive for the transverse simple zero considered here. When either coefficient vanishes, the corresponding direction is governed by higher-order terms in the amplitude.

We integrate over a fixed momentum interval $|q|\leq\Lambda$, where $\Lambda$ sets the width of the window retained around the zero. Shifting the variable by $u\tau$ contributes regular boundary terms, and after subtracting the value at $\tau=0$ the remaining logarithmic contribution is
\begin{equation}
I(\tau)=\int_{-\Lambda}^{\Lambda}
\ln\left(1+\frac{b\tau^2}{a q^2}\right)\,dq .
\label{eq:s_logintegral}
\end{equation}
The logarithm is integrable at $q=0$, but its expansion in powers of $\tau$ is nonuniform in that region. Introducing $\zeta=\sqrt{b/a}|\tau|$, we find $dI/d\zeta=4\arctan(\Lambda/\zeta)$, which approaches $2\pi$ as $\zeta\to0^+$. Together with $I(0)=0$ this yields $I(\tau)=2\pi\sqrt{b/a}|\tau|+O(\tau^2)$.

For a positive momentum normalization $C_k$, the singular part of the return rate is
\begin{equation}
\lambda_{\mathrm{sing}}(\tau)=-2\pi C_k\sqrt{b/a}\,|\tau|+o(|\tau|).
\label{eq:s_cusp}
\end{equation}
The regular background contains the analytic terms, including any contribution linear in $\tau$. The absolute-value term produces the cusp, with a coefficient determined by the local derivatives of the complex amplitude.

For the stored energy the local kernel is proportional to $1-|\mathcal G|^2$. Inserting Eq.~\eqref{eq:s_localnormal} gives ordinary powers of $q$ and $\tau$, multiplied by the smooth initial energy scale, and their momentum integrals remain regular. For the smooth bounded bands used in the main text the energy kernel and its time derivatives are uniformly integrable on the compact momentum interval, so the energy has an ordinary expansion around $t_c$.

This finite-time result relies on the regularity of the band and of the energy weight. A singular initial spectrum, a singular momentum weight, or a nonuniform long-time limit changes the structure of the integral. The long-time-averaged anisotropy response studied in the next appendix involves such a nonuniform limit in the quench parameter.

\section{The anisotropy cusp of the long-time-averaged XY energy}
\label{app:xy_cusp}

We consider the fixed-field scan $g_i=g_f=g$, with $\gamma_i=1$ and $\gamma_f=\gamma$. Define $\delta_k=g-\cos k$ and $s_k=\sin k$, so that the squared initial and final band scales are $M_i^2=\delta_k^2+s_k^2$ and $M_f^2=\delta_k^2+\gamma^2s_k^2$, while the numerator in Eq.~\eqref{eq:xy_overlap} becomes $\delta_k^2+\gamma s_k^2$.

To obtain the transition weight we subtract the squared overlap numerator from $M_i^2M_f^2$. The terms proportional to $\delta_k^4$ and $\gamma^2s_k^4$ cancel, leaving $\delta_k^2s_k^2(1-\gamma)^2$. Hence
\begin{equation}
A(k)=\frac{\delta_k^2s_k^2(1-\gamma)^2}
{(\delta_k^2+s_k^2)(\delta_k^2+\gamma^2s_k^2)}.
\label{eq:s_Axy}
\end{equation}
The factor $(1-\gamma)^2$ follows from the initial value $\gamma_i=1$ and gives the regular asymmetry between positive and negative final anisotropies. The singular dependence arises from the final denominator when $\delta_k$ and $\gamma$ vanish together.

For $|g|<1$ the final spectrum closes at $\gamma=0$ and $k_0=\arccos g$. Writing $s_0=\sqrt{1-g^2}$ and $x=k-k_0$, we have $\delta_k=s_0x+O(x^2)$ and $s_k=s_0+O(x)$, while the initial energy at this momentum is finite and equal to $\epsilon_i(k_0)=2s_0$. The local energy kernel therefore contains the factor $x^2/(x^2+\gamma^2)$.

For fixed nonzero $x$ this factor approaches one as $\gamma\to0$, but within $|x|\sim|\gamma|$ the convergence is nonuniform. The contribution from this shrinking interval follows from
\begin{equation}
\int_{-\Lambda}^{\Lambda}\frac{x^2}{x^2+\gamma^2}\,dx
=2\Lambda-2|\gamma|\arctan\frac{\Lambda}{|\gamma|},
\label{eq:s_localenergy}
\end{equation}
whose leading nonanalytic term is $-\pi|\gamma|$. Restoring $\epsilon_i(k_0)$ and the positive-momentum normalization gives $-s_0|\gamma|$ in the long-time-averaged energy density. A finite cusp therefore survives the momentum integration even though the contributing interval narrows to zero and each individual sector energy remains bounded.

At $\gamma=0$ the transition kernel away from $k_0$ reduces to $s_k^2/M_i^2$, and the long-time-averaged energy is
\begin{equation}
\Delta e_\infty(0)=\frac{1}{\pi}\int_0^\pi
\frac{\sin^2k}{\sqrt{(g-\cos k)^2+\sin^2k}}\,dk .
\label{eq:s_cusp_background}
\end{equation}
The isolated point $k_0$ has zero measure. Apart from the prefactor $(1-\gamma)^2$, the dependence on $\gamma$ is even, and this prefactor supplies the regular coefficient $B(g)=-2\Delta e_\infty(0)$. Combining the regular and nonanalytic terms gives $\Delta e_\infty(\gamma)=\Delta e_\infty(0)+B(g)\gamma-s_0|\gamma|+o(|\gamma|)$.

Taking the one-sided derivatives yields Eq.~\eqref{eq:xy_cusp}. The magnitude of the slope jump is $2s_0$ and comes from the closing mode, while the mean of the two limiting slopes is the regular value $B(g)$ supplied by the initial anisotropy.

For $|g|>1$ the quantity $\delta_k$ is nonzero throughout the momentum interval, and the denominator of Eq.~\eqref{eq:s_Axy} remains bounded away from zero near $\gamma=0$, which gives a smooth long-time-averaged energy, as shown for $g=1.2$. At $|g|=1$ the closing momentum lies at an endpoint of the Brillouin zone, and the local scaling differs from the interior-momentum expansion used above.

\section{Exact diagonalization and the long-time energy comparison}
\label{app:ed}

We evaluate the interacting Hamiltonian of Eq.~\eqref{eq:nnn_model} with periodic boundaries and $J=1$. Lattice translation and global spin flip commute with both the initial and the final Hamiltonian. For the parameters studied here the initial ground state belongs to the zero-momentum, even spin-flip sector, so the evolution is confined to this symmetry block.

A configuration in the $\sigma^z$ basis is represented by a bit string. Its lattice translations and globally flipped partners form an orbit $\mathcal O_a$, and the normalized basis state is $|\mathcal O_a\rangle=|\mathcal O_a|^{-1/2}\sum_{s\in\mathcal O_a}|s\rangle$, where $|\mathcal O_a|$ is the number of configurations in the orbit. The nearest- and next-nearest-neighbor Ising energies are constant along each orbit, so their diagonal matrix elements are evaluated from one representative. The full and reduced dimensions are listed in Table~\ref{tab:dims}.

The transverse operator $X=\sum_j\sigma_j^x$ acts by flipping one bit. Let $n_{b\leftarrow a}$ count the single-bit flips of a representative of orbit $a$ that reach orbit $b$. The matrix element is
\begin{equation}
\langle\mathcal O_b|X|\mathcal O_a\rangle
=n_{b\leftarrow a}\sqrt{\frac{|\mathcal O_a|}{|\mathcal O_b|}}.
\label{eq:s_orbitmatrix}
\end{equation}
Before normalization the number of contributing processes is $|\mathcal O_a|n_{b\leftarrow a}$, and dividing by the normalization factors of both orbit states gives the square-root factor of Eq.~\eqref{eq:s_orbitmatrix}.

\begin{table}[tbp]
\caption{Hilbert-space dimensions for the interacting chain and the zero-momentum, even spin-flip blocks preserved by the quench.}
\label{tab:dims}
\begin{ruledtabular}
\begin{tabular}{ccc}
$N$ & Full dimension $2^N$ & Reduced dimension\\
8 & 256 & 20\\
10 & 1024 & 56\\
12 & 4096 & 180\\
14 & 16384 & 596
\end{tabular}
\end{ruledtabular}
\end{table}

We diagonalize $H_f|m_f\rangle=E_m^f|m_f\rangle$, where $|m_f\rangle$ and $E_m^f$ are the eigenstates and eigenvalues of the final Hamiltonian, and define the overlaps $c_m=\langle m_f|\psi_0\rangle$. The state at time $t$ is
\begin{equation}
|\psi(t)\rangle=\sum_m c_m e^{-iE_m^ft}|m_f\rangle,
\label{eq:s_evolution}
\end{equation}
with the sum running over every eigenstate in the accessible symmetry sector. We evaluate the phases directly and calculate $\mathcal G(t)=\sum_m|c_m|^2e^{-iE_m^ft}$. The stored energy follows from Eq.~\eqref{eq:stored_energy}, and the power is evaluated through the commutator in Eq.~\eqref{eq:interacting_power}. Together with $A_N=1-|\mathcal G|^2$, this procedure gives all four quantities of Fig.~\ref{fig:nnn_dynamics} from a common time-evolved state.

The return-rate markers in Fig.~\ref{fig:nnn_dynamics}(a) identify the dominant local maxima in the displayed interval. We bracket these peaks on a grid of $1001$ points over $0.35\leq Jt\leq1.05$ and then refine their positions by bounded optimization of the spectral expectation value. The field scan uses $61$ values over $0.05\leq g_f\leq0.50$ and $601$ times over $0.25\leq Jt\leq1.25$, with peaks selected inside $0.35\leq Jt\leq1.15$. The power decomposition of Fig.~\ref{fig:nnn_dynamics}(c) uses $801$ times over $0.48\leq Jt\leq0.92$. We obtain $\dot A_N=-2\operatorname{Re}(\mathcal G^*\dot{\mathcal G})$ by differentiating the spectral phases, and we calculate the redistribution contribution from the difference between the total and population powers.

For a nondegenerate final spectrum the long-time average removes the terms with unequal final energies, and the resulting stored-energy density is
\begin{equation}
\overline{\Delta e}_N=\frac{1}{N}\left[
\sum_m|c_m|^2\langle m_f|H_i|m_f\rangle-E_0^i\right].
\label{eq:s_diag_energy}
\end{equation}
This diagonal-ensemble expression evaluates the infinite-time average directly, including the effects of all populated final eigenstates, and the oscillations and recurrences of a finite trajectory are averaged through their spectral phases.

When the final spectrum has exact degeneracies, coherences within a degenerate energy subspace survive the time average. Let $\Pi_\mu^f$ project onto a distinct final energy, labeled by $\mu$. The basis-independent result is
\begin{equation}
\overline{\Delta e}_N=\frac{1}{N}\left[
\sum_\mu\langle\psi_0|\Pi_\mu^f H_i\Pi_\mu^f|\psi_0\rangle-E_0^i\right].
\label{eq:s_projectoraverage}
\end{equation}
For one-dimensional energy eigenspaces this reduces to Eq.~\eqref{eq:s_diag_energy}. The parameter dependence of the long-time-averaged observable follows that of the relevant eigenspaces and of the coherences retained by the average, so it is smooth on intervals where these quantities vary smoothly. At crossings the spectral projectors determine the limiting behavior, while finite-time expectation values for a finite matrix Hamiltonian remain regular under smooth parameter changes.

\begin{figure}[tb]
\centering
\includegraphics[width=\columnwidth]{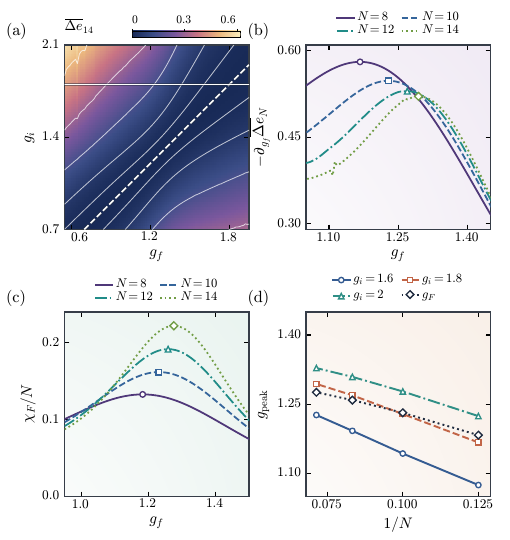}
\caption{Finite-quench response and equilibrium fidelity susceptibility at $\kappa=0.2$. (a) Long-time-averaged energy landscape $\overline{\Delta e}_{14}$ over the preparation field $g_i$ and the final field $g_f$, with the diagonal $g_f=g_i$ marked. (b) Charging-response derivative $-\partial_{g_f}\overline{\Delta e}_N$ at $g_i=1.8$ for four sizes. (c) Equilibrium fidelity susceptibility per site $\chi_F/N$. (d) Peak field $g_{\mathrm{peak}}$ against inverse size for three preparation fields, compared with the peak $g_F$ of the fidelity susceptibility.}
\label{fig:int_steady}
\end{figure}

Figure~\ref{fig:int_steady}(a) scans both preparation and final fields at $N=14$ and $\kappa=0.2$. We use $71$ initial fields over $0.70\leq g_i\leq2.10$ and $101$ final fields over $0.55\leq g_f\leq1.95$. For the peak comparison of Fig.~\ref{fig:int_steady}(b) through Fig.~\ref{fig:int_steady}(d) we refine the final-field step to $0.0025$ over $0.95\leq g_f\leq1.50$. The derivative is evaluated by centered finite differences in the interior and by second-order one-sided differences at the endpoints, and Fig.~\ref{fig:int_steady}(b) displays the broad-response interval $1.05\leq g_f\leq1.45$.

For comparison we calculate the equilibrium ground-state fidelity susceptibility at each scanned field. The ground-state-overlap approach to quantum phase transitions was developed by Zanardi and Paunkovi\'c \cite{ZanardiPaunkovic2006}, and fidelity susceptibility together with its relation to the dynamical structure factor was formulated by You, Li, and Gu \cite{YouLiGu2007}. A broader review is given in Ref.~\cite{Gu2010Fidelity}. We adopt the amplitude-fidelity convention
\begin{equation}
\mathcal F(g,\delta g)
=
\bigl|\langle 0(g)|0(g+\delta g)\rangle\bigr|
=
1-\frac{(\delta g)^2}{2}\chi_F(g)+O((\delta g)^3),
\label{eq:s_fidelity_definition}
\end{equation}
in which $\mathcal F$ is the overlap between the ground states at two nearby fields and $\chi_F$ is the fidelity susceptibility. For a nondegenerate ground state, standard perturbation theory then gives
\begin{equation}
\frac{\chi_F}{N}=\frac{1}{N}\sum_{n>0}
\frac{|\langle n|V|0\rangle|^2}{(E_n-E_0)^2},
\qquad
V=\partial_g H.
\label{eq:s_fidelity}
\end{equation}
When the squared fidelity $\mathcal F^2$ is used instead, its quadratic coefficient is $\chi_F$ rather than $\chi_F/2$, and Eq.~\eqref{eq:s_fidelity} itself is unchanged. In the transverse-field scan and in the units used numerically, $V=-\sum_j\sigma_j^x$. We evaluate Eq.~\eqref{eq:s_fidelity} from the complete eigenbasis of the same symmetry block used for the charging calculation. The initial preparation field does not enter this equilibrium quantity, since $\chi_F$ depends only on the ground state and the excitation spectrum of the Hamiltonian at the field being scanned.

Figure~\ref{fig:int_steady}(a) first shows the full finite-quench landscape. The white diagonal $g_f=g_i$ is dark because the state is not changed there and the stored energy vanishes identically. Moving away from the diagonal populates excited states of the initial Hamiltonian and increases the long-time-averaged energy, but the contours differ visibly above and below the diagonal. This asymmetry is physical, since exchanging $g_i$ and $g_f$ changes both the prepared ground state and the Hamiltonian with respect to which the battery energy is measured. The horizontal line at $g_i=1.8$ therefore selects one preparation-specific trajectory through a genuinely two-parameter response surface, and it does not represent a property of the final Hamiltonian alone. The broad variation away from the diagonal also shows that finite-quench charging samples a substantial portion of the spectrum rather than only the lowest gap, which is why the response surface can remain smooth even when the scan crosses the finite-size remnant of an equilibrium critical region.

Figure~\ref{fig:int_steady}(b) differentiates that trajectory with respect to $g_f$. The broad maximum identifies the field interval in which the finite-quench energy changes most rapidly. With increasing size the location of this maximum drifts while its height decreases over the sizes shown, which differs from what one would expect if the curve were simply reproducing an equilibrium critical susceptibility. The derivative contains the response of a finite-amplitude diagonal ensemble, and it depends on the overlaps of the chosen initial ground state with many final eigenstates and on their energies measured by $H_i$. Its maximum therefore marks the point of fastest change of the charging protocol rather than a thermodynamic singularity. The broadness of the peak is meaningful as well, since several excitation channels contribute over a finite field interval and no single low-energy mode controls the observable.

Figure~\ref{fig:int_steady}(c) provides the equilibrium comparison. The fidelity susceptibility develops a sharper and higher peak with increasing $N$, because every transition is weighted by $|V_{n0}|^2/\omega_{n0}^2$ and the extra inverse power of the gap strongly magnifies the lowest-energy excitations. The finite-quench energy derivative of Fig.~\ref{fig:int_steady}(b) contains a broader mixture of excited states and retains explicit information about the preparation field. The opposite size evolution of the two peak heights therefore has a simple spectral origin, and it cautions against identifying the largest finite-quench charging response with the equilibrium fidelity peak. The same comparison explains why the two quantities can peak in a similar field region without being equivalent. Both are sensitive to spectral rearrangement, but the squared energy denominator in $\chi_F$ selectively amplifies the softest excitations, whereas the charging observable retains appreciable weight from higher-energy states.

Figure~\ref{fig:int_steady}(d) makes the distinction direct by comparing peak positions. Changing $g_i$ from $1.6$ to $2.0$ shifts the charging-response sequence even though the set of final Hamiltonians is the same, while the fidelity-susceptibility peak $g_F$ is independent of $g_i$ because it is an equilibrium property of $H(g_f)$. The three charging sequences do not collapse onto $g_F$ over the available sizes, so their preparation dependence is a defining feature of a finite quench rather than a finite-size plotting artifact. If the charging maximum were governed by the same equilibrium spectral sum as $\chi_F$, changing $g_i$ would not generate three distinct peak sequences at a fixed family of final Hamiltonians. The separation visible in Fig.~\ref{fig:int_steady}(d) is therefore a direct diagnostic that the finite-quench response carries memory of the preparation.

The four parts of Fig.~\ref{fig:int_steady} together explain why the finite-quench derivative and the fidelity susceptibility should not be used interchangeably. The former measures the energy injected by a specific pair $(H_i,H_f)$ and samples a finite population of many final eigenstates, while the latter measures the infinitesimal change of an equilibrium ground state and overweights the lowest gaps. The appropriate equilibrium quantity for the battery emerges only in the controlled weak-quench expansion derived in Appendix~\ref{app:small_quench}, where the leading energy is governed by the static susceptibility with a single inverse-gap denominator.

\section{Small-quench response without integrability}
\label{app:small_quench}

This appendix derives the weak-quench result used in Sec.~\ref{sec:weak_quench} and makes explicit its relation to both the ordinary static susceptibility and the fidelity susceptibility. The derivation is carried out at fixed finite size. We assume that the initial ground state is nondegenerate and separated from the excited states by nonzero finite-size gaps, and we note below that excited-state degeneracies do not change the final spectral formula when they are treated with projectors or degenerate perturbation theory.

Let $H(g)$ be a smooth family of Hamiltonians. At the initial parameter $g$ we write
\begin{equation}
H_i=H(g),\qquad
H_i|n\rangle=E_n|n\rangle,
\end{equation}
with ground state $|0\rangle$, ground-state energy $E_0$, and excitation energies
\begin{equation}
\omega_{n0}=E_n-E_0>0 \qquad (n>0).
\label{eq:s_gap_definition}
\end{equation}
After a sudden quench $g\rightarrow g+\delta g$,
\begin{equation}
H_f
=
H_i+\delta g\,V+\frac{(\delta g)^2}{2}W+O((\delta g)^3),
\label{eq:s_quench}
\end{equation}
where
\begin{equation}
V=\partial_gH,\qquad W=\partial_g^2H
\end{equation}
are evaluated at the initial parameter. The leading charging response depends only on $V$, and the operator $W$ first enters the higher-order correction.

For transparency we first suppose that the relevant levels are nondegenerate. First-order perturbation theory gives the component of the final ground state orthogonal to $|0\rangle$,
\begin{equation}
|0_f\rangle
=
\mathcal N_0
\left(
|0\rangle
-
\delta g\sum_{n>0}\frac{V_{n0}}{\omega_{n0}}|n\rangle
\right)
+O((\delta g)^2)_{\perp},
\label{eq:s_groundmixing}
\end{equation}
where $V_{n0}=\langle n|V|0\rangle$ and the normalization factor is
\begin{equation}
\mathcal N_0
=
1-\frac{(\delta g)^2}{2}
\sum_{n>0}\frac{|V_{n0}|^2}{\omega_{n0}^2}
+O((\delta g)^3).
\label{eq:s_groundnorm}
\end{equation}
The symbol $O((\delta g)^2)_{\perp}$ denotes second-order components orthogonal to the unperturbed ground state.

Likewise, the final eigenstate continuously connected to $|n\rangle$ contains the ground-state component
\begin{equation}
|n_f\rangle
=
|n\rangle
+
\delta g\frac{V_{0n}}{\omega_{n0}}|0\rangle
+O(\delta g)_{\{m\neq0,n\}}
+O((\delta g)^2).
\label{eq:s_excitedmixing}
\end{equation}
Consequently,
\begin{align}
p_0&\equiv|\langle0_f|0\rangle|^2
=1-(\delta g)^2\sum_{n>0}\frac{|V_{n0}|^2}{\omega_{n0}^2}
+O((\delta g)^3),
\label{eq:s_p0}
\\
p_n&\equiv|\langle n_f|0\rangle|^2
=(\delta g)^2\frac{|V_{n0}|^2}{\omega_{n0}^2}
+O((\delta g)^3),\qquad n>0.
\label{eq:s_pn}
\end{align}
The loss of final-ground-state occupation is therefore exactly the total excited-state occupation to this order.

The battery energy is measured with the initial Hamiltonian. For a nondegenerate final spectrum the infinite-time average removes coherences between distinct final energies and gives
\begin{equation}
\overline{\rho}
=
\sum_m p_m |m_f\rangle\langle m_f|,
\qquad
\overline{\Delta E}
=
\operatorname{Tr}(H_i\overline{\rho})-E_0,
\label{eq:s_long_time_state}
\end{equation}
where $\overline{\rho}$ is the time-averaged density matrix.

First, because $H_i$ is diagonal in the unperturbed basis, the linear correction to $\langle0_f|H_i|0_f\rangle$ vanishes. Using Eqs.~\eqref{eq:s_groundmixing} and \eqref{eq:s_groundnorm},
\begin{align}
\langle0_f|H_i|0_f\rangle-E_0
&=(\delta g)^2
\sum_{n>0}\frac{|V_{n0}|^2}{\omega_{n0}^2}(E_n-E_0)
+O((\delta g)^3)
\\
&=(\delta g)^2
\sum_{n>0}\frac{|V_{n0}|^2}{\omega_{n0}}
+O((\delta g)^3).
\label{eq:s_ground_contribution}
\end{align}
Since $p_0=1+O((\delta g)^2)$, multiplying this expression by $p_0$ changes it only at fourth order, which is irrelevant at the order retained.

Second, the final excited-state probabilities are already of order $(\delta g)^2$, so only the zeroth-order excitation energy measured by $H_i$ is needed,
\begin{equation}
\langle n_f|H_i|n_f\rangle-E_0
=
\omega_{n0}+O((\delta g)^2).
\label{eq:s_excited_energy}
\end{equation}
There is no term linear in $\delta g$, because $H_i$ is diagonal in its own eigenbasis. Therefore
\begin{align}
&\sum_{n>0}p_n
\bigl(\langle n_f|H_i|n_f\rangle-E_0\bigr)
\nonumber\\
&\qquad=(\delta g)^2
\sum_{n>0}\frac{|V_{n0}|^2}{\omega_{n0}}
+O((\delta g)^3).
\label{eq:s_excited_contribution}
\end{align}
Adding Eqs.~\eqref{eq:s_ground_contribution} and \eqref{eq:s_excited_contribution} gives
\begin{equation}
\overline{\Delta E}
=
2(\delta g)^2
\sum_{n>0}\frac{|V_{n0}|^2}{\omega_{n0}}
+O((\delta g)^3)
\label{eq:s_DEbar}
\end{equation}
and therefore
\begin{equation}
\overline{\Delta e}_N
=
(\delta g)^2\chi_V^{(N)}(g)
+O((\delta g)^3),
\label{eq:s_theorem}
\end{equation}
with
\begin{equation}
\chi_V^{(N)}(g)
=
\frac{2}{N}\sum_{n>0}\frac{|V_{n0}|^2}{\omega_{n0}}.
\label{eq:s_chi}
\end{equation}
The factor of two in Eq.~\eqref{eq:s_DEbar} is not an additional convention. One identical spectral sum comes from the deformation of the final ground state when its energy is evaluated with $H_i$, and the other comes from the small population transferred to the final excited states. A direct two-level expansion gives the same factor.

A basis-independent form makes the structure transparent. Let $Q=1-|0\rangle\langle0|$ be the projector onto the excited subspace and define the reduced resolvent
\begin{equation}
R=Q(H_i-E_0)^{-1}Q.
\end{equation}
Then
\begin{equation}
\chi_V^{(N)}
=
\frac{2}{N}\langle0|VRV|0\rangle.
\label{eq:s_resolvent_chi}
\end{equation}
This form remains well defined in the presence of degeneracies inside the excited subspace, provided the ground state itself is nondegenerate.

For a Hamiltonian linear in the control parameter,
\begin{equation}
H(g)=H_0+gV,
\end{equation}
Hellmann--Feynman theory gives
\begin{equation}
\frac{dE_0}{dg}=\langle0|V|0\rangle,
\end{equation}
and differentiating once more yields
\begin{equation}
\frac{d^2E_0}{dg^2}
=
-2\sum_{n>0}\frac{|V_{n0}|^2}{\omega_{n0}}.
\label{eq:s_curvature}
\end{equation}
Thus
\begin{equation}
\chi_V^{(N)}
=
-\frac{1}{N}\frac{d^2E_0}{dg^2},
\qquad
\overline{\Delta e}_N
=
-\frac{(\delta g)^2}{N}\frac{d^2E_0}{dg^2}
+O((\delta g)^3).
\label{eq:s_chi_curvature}
\end{equation}
This sign is easiest to interpret in terms of the generalized force $F_g=-\partial_gH=-V$. Its zero-temperature static response to $g$ is $N^{-1}d\langle F_g\rangle/dg=-N^{-1}d^2E_0/dg^2=\chi_V^{(N)}$. For the transverse-field form $H=H_0-gM$ one has $V=-M$, and Eq.~\eqref{eq:s_chi_curvature} is precisely the static transverse susceptibility per site, $N^{-1}d\langle M\rangle/dg$, in the dimensionless units used here. Because it is a ground-state curvature, this quantity also carries the information that classifies a quantum phase transition, whose order is fixed by the lowest derivative of the ground-state energy that becomes singular. A continuous transition makes the curvature diverge at the critical point, whereas a first-order transition makes the energy slope jump and leaves a curvature spike of finite weight.

If $H(g)$ depends nonlinearly on $g$, the ground-state curvature contains an additional term,
\begin{equation}
\frac{d^2E_0}{dg^2}
=
\langle0|\partial_g^2H|0\rangle
-2\sum_{n>0}
\frac{|\langle n|\partial_gH|0\rangle|^2}{\omega_{n0}}.
\label{eq:s_nonlinear}
\end{equation}
The spectral sum in Eq.~\eqref{eq:s_chi} still controls the leading charging energy, but it can no longer be identified with $-N^{-1}d^2E_0/dg^2$ without subtracting the explicit $\langle\partial_g^2H\rangle$ contribution.

The distinction from the fidelity susceptibility can now be stated without ambiguity. With the fidelity convention of Eq.~\eqref{eq:s_fidelity_definition}, perturbation theory gives \cite{YouLiGu2007,Gu2010Fidelity}
\begin{equation}
\chi_F
=
\sum_{n>0}\frac{|V_{n0}|^2}{\omega_{n0}^2}
=
\langle0|VR^2V|0\rangle.
\label{eq:s_chiF_again}
\end{equation}
The weak-quench charging coefficient and the fidelity susceptibility therefore involve the same matrix elements but different moments of the excitation spectrum,
\begin{equation}
\chi_V^{(N)}
=\frac{2}{N}\sum_{n>0}\frac{|V_{n0}|^2}{\omega_{n0}},
\qquad
\frac{\chi_F}{N}
=\frac{1}{N}\sum_{n>0}\frac{|V_{n0}|^2}{\omega_{n0}^2}.
\label{eq:s_chi_compare}
\end{equation}
The extra inverse gap in $\chi_F$ makes it more infrared sensitive, which is the precise reason why the fidelity-susceptibility peak in Fig.~\ref{fig:int_steady}(c) can be sharper and can scale differently from the finite-quench charging response.

The connection can also be written in the spectral language emphasized by You, Li, and Gu \cite{YouLiGu2007}. Define the zero-temperature spectral density of the driving operator, with the normalization used here, as
\begin{equation}
\mathcal S_V(\omega)
=
\frac{1}{N}\sum_{n>0}|V_{n0}|^2
\delta(\omega-\omega_{n0}),
\label{eq:s_spectral_density}
\end{equation}
where $\omega$ is the frequency variable and $\delta(\cdot)$ is the Dirac delta function. Then
\begin{equation}
\chi_V^{(N)}
=2\int_0^\infty d\omega\,
\frac{\mathcal S_V(\omega)}{\omega},
\qquad
\frac{\chi_F}{N}
=\int_0^\infty d\omega\,
\frac{\mathcal S_V(\omega)}{\omega^2}.
\label{eq:s_spectral_moments}
\end{equation}
Equation~\eqref{eq:s_spectral_moments} shows directly that the two observables are related without being interchangeable. The charging coefficient is the first inverse-frequency moment of the same spectral density whose second inverse-frequency moment gives the fidelity susceptibility, and a soft mode is weighted by one additional power of $1/\omega$ in $\chi_F$.

If the final Hamiltonian has exact degeneracies, the time average must be written with projectors $\Pi_\mu^f$ onto distinct final energies,
\begin{equation}
\overline{\rho}
=
\sum_\mu \Pi_\mu^f |0\rangle\langle0|\Pi_\mu^f,
\label{eq:s_long_time_projectors}
\end{equation}
which is the density-matrix counterpart of Eq.~\eqref{eq:s_projectoraverage}. Degenerate perturbation theory is then performed within each spectral subspace. For a nondegenerate initial ground state and no level crossing involving that ground state over the infinitesimal quench, the leading coefficient remains Eq.~\eqref{eq:s_chi}, and the resolvent expression of Eq.~\eqref{eq:s_resolvent_chi} is independent of the basis chosen inside a degenerate excited manifold. If the ground state itself becomes exactly degenerate, ordinary nondegenerate perturbation theory no longer applies and the weak-quench limit must be formulated within the ground-state manifold.

Collecting the two contributions,
\begin{equation}
\lim_{\delta g\rightarrow0}
\frac{\overline{\Delta e}_N}{(\delta g)^2}
=
\chi_V^{(N)}
\label{eq:s_limit}
\end{equation}
is a fixed-$N$ statement. Near a thermodynamic critical point the smallest finite-size gap decreases with $N$, so the interval of $\delta g$ over which first-order eigenstate perturbation theory is accurate can shrink. A useful sufficient condition is that the mixing amplitudes remain small,
\begin{equation}
|\delta g|\,
\frac{|V_{n0}|}{\omega_{n0}}\ll1
\end{equation}
for the low-energy states carrying appreciable spectral weight. One should therefore establish the $\delta g\rightarrow0$ convergence at each finite size before extracting the size dependence of $\chi_V^{(N)}$, which is the order of limits tested numerically in Fig.~\ref{fig:small_quench}.

%% file: references.bib
@PREAMBLE{
 "\providecommand{\noopsort}[1]{}" 
 # "\providecommand{\singleletter}[1]{#1}%" 
}

@book{Sachdev2011,
  title     = {Quantum Phase Transitions},
  author    = {Sachdev, Subir},
  edition   = {2},
  publisher = {Cambridge University Press},
  address   = {Cambridge},
  year      = {2011}
}

@article{RevModPhys.82.3045,
  title = {Colloquium: Topological insulators},
  author = {Hasan, M. Z. and Kane, C. L.},
  journal = {Rev. Mod. Phys.},
  volume = {82},
  issue = {4},
  pages = {3045--3067},
  numpages = {0},
  year = {2010},
  month = {Nov},
  publisher = {American Physical Society},
  doi = {10.1103/RevModPhys.82.3045},
  url = {https://link.aps.org/doi/10.1103/RevModPhys.82.3045}
}

@article{RevModPhys.83.1057,
  title = {Topological insulators and superconductors},
  author = {Qi, Xiao-Liang and Zhang, Shou-Cheng},
  journal = {Rev. Mod. Phys.},
  volume = {83},
  issue = {4},
  pages = {1057--1110},
  numpages = {0},
  year = {2011},
  month = {Oct},
  publisher = {American Physical Society},
  doi = {10.1103/RevModPhys.83.1057},
  url = {https://link.aps.org/doi/10.1103/RevModPhys.83.1057}
}

@article{Ryu_2010,
doi = {10.1088/1367-2630/12/6/065010},
url = {https://doi.org/10.1088/1367-2630/12/6/065010},
year = {2010},
month = {jun},
publisher = {},
volume = {12},
number = {6},
pages = {065010},
author = {Ryu, Shinsei and Schnyder, Andreas P and Furusaki, Akira and Ludwig, Andreas W W},
title = {Topological insulators and superconductors: tenfold way and dimensional hierarchy},
journal = {New Journal of Physics}
}

@article{RevModPhys.88.035005,
  title = {Classification of topological quantum matter with symmetries},
  author = {Chiu, Ching-Kai and Teo, Jeffrey C. Y. and Schnyder, Andreas P. and Ryu, Shinsei},
  journal = {Rev. Mod. Phys.},
  volume = {88},
  issue = {3},
  pages = {035005},
  numpages = {63},
  year = {2016},
  month = {Aug},
  publisher = {American Physical Society},
  doi = {10.1103/RevModPhys.88.035005},
  url = {https://link.aps.org/doi/10.1103/RevModPhys.88.035005}
}

@article{RevModPhys.69.315,
  title = {Continuous quantum phase transitions},
  author = {Sondhi, S. L. and Girvin, S. M. and Carini, J. P. and Shahar, D.},
  journal = {Rev. Mod. Phys.},
  volume = {69},
  issue = {1},
  pages = {315--333},
  numpages = {0},
  year = {1997},
  month = {Jan},
  publisher = {American Physical Society},
  doi = {10.1103/RevModPhys.69.315},
  url = {https://link.aps.org/doi/10.1103/RevModPhys.69.315}
}

@article{MatthiasVojta_2003,
doi = {10.1088/0034-4885/66/12/R01},
url = {https://doi.org/10.1088/0034-4885/66/12/R01},
year = {2003},
month = {nov},
publisher = {},
volume = {66},
number = {12},
pages = {2069},
author = {Matthias Vojta},
title = {Quantum phase transitions},
journal = {Reports on Progress in Physics}
}

@article{PhysRevE.87.042123,
  title = {Entanglement boost for extractable work from ensembles of quantum batteries},
  author = {Alicki, Robert and Fannes, Mark},
  journal = {Phys. Rev. E},
  volume = {87},
  issue = {4},
  pages = {042123},
  numpages = {4},
  year = {2013},
  month = {Apr},
  publisher = {American Physical Society},
  doi = {10.1103/PhysRevE.87.042123},
  url = {https://link.aps.org/doi/10.1103/PhysRevE.87.042123}
}

@Inbook{Campaioli2018,
author="Campaioli, Francesco
and Pollock, Felix A.
and Vinjanampathy, Sai",
editor="Binder, Felix
and Correa, Luis A.
and Gogolin, Christian
and Anders, Janet
and Adesso, Gerardo",
title="Quantum Batteries",
bookTitle="Thermodynamics in the Quantum Regime: Fundamental Aspects and New Directions",
year="2018",
publisher="Springer International Publishing",
address="Cham",
pages="207--225",
isbn="978-3-319-99046-0",
doi="10.1007/978-3-319-99046-0_8",
url="https://doi.org/10.1007/978-3-319-99046-0_8"
}

@article{Binder2015,
doi = {10.1088/1367-2630/17/7/075015},
url = {https://doi.org/10.1088/1367-2630/17/7/075015},
year = {2015},
month = {jul},
publisher = {IOP Publishing},
volume = {17},
number = {7},
pages = {075015},
author = {Binder, Felix C and Vinjanampathy, Sai and Modi, Kavan and Goold, John},
title = {Quantacell: powerful charging of quantum batteries},
journal = {New Journal of Physics}
}

@article{PhysRevA.107.023725,
  title = {Superconducting transmon qubit-resonator quantum battery},
  author = {Dou, Fu-Quan and Yang, Fang-Mei},
  journal = {Phys. Rev. A},
  volume = {107},
  issue = {2},
  pages = {023725},
  numpages = {10},
  year = {2023},
  month = {Feb},
  publisher = {American Physical Society},
  doi = {10.1103/PhysRevA.107.023725},
  url = {https://link.aps.org/doi/10.1103/PhysRevA.107.023725}
}

@article{PhysRevLett.122.210601,
  title = {Dissipative Charging of a Quantum Battery},
  author = {Barra, Felipe},
  journal = {Phys. Rev. Lett.},
  volume = {122},
  issue = {21},
  pages = {210601},
  numpages = {6},
  year = {2019},
  month = {May},
  publisher = {American Physical Society},
  doi = {10.1103/PhysRevLett.122.210601},
  url = {https://link.aps.org/doi/10.1103/PhysRevLett.122.210601}
}

@article{PhysRevB.99.035421,
  title = {Charger-mediated energy transfer for quantum batteries: An open-system approach},
  author = {Farina, Donato and Andolina, Gian Marcello and Mari, Andrea and Polini, Marco and Giovannetti, Vittorio},
  journal = {Phys. Rev. B},
  volume = {99},
  issue = {3},
  pages = {035421},
  numpages = {15},
  year = {2019},
  month = {Jan},
  publisher = {American Physical Society},
  doi = {10.1103/PhysRevB.99.035421},
  url = {https://link.aps.org/doi/10.1103/PhysRevB.99.035421}
}

@article{PhysRevLett.122.047702,
  title = {Extractable Work, the Role of Correlations, and Asymptotic Freedom in Quantum Batteries},
  author = {Andolina, Gian Marcello and Keck, Maximilian and Mari, Andrea and Campisi, Michele and Giovannetti, Vittorio and Polini, Marco},
  journal = {Phys. Rev. Lett.},
  volume = {122},
  issue = {4},
  pages = {047702},
  numpages = {5},
  year = {2019},
  month = {Feb},
  publisher = {American Physical Society},
  doi = {10.1103/PhysRevLett.122.047702},
  url = {https://link.aps.org/doi/10.1103/PhysRevLett.122.047702}
}

@article{PhysRevA.109.042207,
  title = {Quantum battery with non-Hermitian charging},
  author = {Konar, Tanoy Kanti and Lakkaraju, Leela Ganesh Chandra and Sen (De), Aditi},
  journal = {Phys. Rev. A},
  volume = {109},
  issue = {4},
  pages = {042207},
  numpages = {12},
  year = {2024},
  month = {Apr},
  publisher = {American Physical Society},
  doi = {10.1103/PhysRevA.109.042207},
  url = {https://link.aps.org/doi/10.1103/PhysRevA.109.042207}
}

@article{d9k1-75d4,
  title = {Self-Discharging Mitigated Quantum Battery},
  author = {Song, Wan-Lu and Wang, Ji-Ling and Zhou, Bin and Yang, Wan-Li and An, Jun-Hong},
  journal = {Phys. Rev. Lett.},
  volume = {135},
  issue = {2},
  pages = {020405},
  numpages = {8},
  year = {2025},
  month = {Jul},
  publisher = {American Physical Society},
  doi = {10.1103/d9k1-75d4},
  url = {https://link.aps.org/doi/10.1103/d9k1-75d4}
}

@article{6kwv-z6fx,
  title = {Reliable quantum advantage in quantum battery charging},
  author = {Rinaldi, Davide and Filip, Radim and Gerace, Dario and Guarnieri, Giacomo},
  journal = {Phys. Rev. A},
  volume = {112},
  issue = {1},
  pages = {012205},
  numpages = {11},
  year = {2025},
  month = {Jul},
  publisher = {American Physical Society},
  doi = {10.1103/6kwv-z6fx},
  url = {https://link.aps.org/doi/10.1103/6kwv-z6fx}
}

@article{PhysRevLett.128.140501,
  title = {Quantum Charging Advantage Cannot Be Extensive without Global Operations},
  author = {Gyhm, Ju-Yeon and \ifmmode \check{S}\else \v{S}\fi{}afr\'anek, Dominik and Rosa, Dario},
  journal = {Phys. Rev. Lett.},
  volume = {128},
  issue = {14},
  pages = {140501},
  numpages = {6},
  year = {2022},
  month = {Apr},
  publisher = {American Physical Society},
  doi = {10.1103/PhysRevLett.128.140501},
  url = {https://link.aps.org/doi/10.1103/PhysRevLett.128.140501}
}

@article{RevModPhys.96.031001,
  title = {Colloquium: Quantum batteries},
  author = {Campaioli, Francesco and Gherardini, Stefano and Quach, James Q. and Polini, Marco and Andolina, Gian Marcello},
  journal = {Rev. Mod. Phys.},
  volume = {96},
  issue = {3},
  pages = {031001},
  numpages = {30},
  year = {2024},
  month = {Jul},
  publisher = {American Physical Society},
  doi = {10.1103/RevModPhys.96.031001},
  url = {https://link.aps.org/doi/10.1103/RevModPhys.96.031001}
}

@article{Ferraro2026,
  author    = {Ferraro, Dario and Cavaliere, Fabio and Genoni, Marco G. and Benenti, Giuliano and Sassetti, Maura},
  title     = {Opportunities and challenges of quantum batteries},
  journal   = {Nature Reviews Physics},
  year      = {2026},
  volume    = {8},
  number    = {2},
  pages     = {115--127},
  doi       = {10.1038/s42254-025-00906-5},
  url       = {https://doi.org/10.1038/s42254-025-00906-5},
  issn      = {2522-5820}
}

@article{Campaioli2017,
  title = {Enhancing the Charging Power of Quantum Batteries},
  author = {Campaioli, Francesco and Pollock, Felix A. and Binder, Felix C. and C\'eleri, Lucas and Goold, John and Vinjanampathy, Sai and Modi, Kavan},
  journal = {Phys. Rev. Lett.},
  volume = {118},
  issue = {15},
  pages = {150601},
  numpages = {6},
  year = {2017},
  month = {Apr},
  publisher = {American Physical Society},
  doi = {10.1103/PhysRevLett.118.150601},
  url = {https://link.aps.org/doi/10.1103/PhysRevLett.118.150601}
}

@article{Ferraro2018,
  title = {High-Power Collective Charging of a Solid-State Quantum Battery},
  author = {Ferraro, Dario and Campisi, Michele and Andolina, Gian Marcello and Pellegrini, Vittorio and Polini, Marco},
  journal = {Phys. Rev. Lett.},
  volume = {120},
  issue = {11},
  pages = {117702},
  numpages = {6},
  year = {2018},
  month = {Mar},
  publisher = {American Physical Society},
  doi = {10.1103/PhysRevLett.120.117702},
  url = {https://link.aps.org/doi/10.1103/PhysRevLett.120.117702}
}

@article{Andolina2019PRB,
  title = {Quantum versus classical many-body batteries},
  author = {Andolina, Gian Marcello and Keck, Maximilian and Mari, Andrea and Giovannetti, Vittorio and Polini, Marco},
  journal = {Phys. Rev. B},
  volume = {99},
  issue = {20},
  pages = {205437},
  numpages = {7},
  year = {2019},
  month = {May},
  publisher = {American Physical Society},
  doi = {10.1103/PhysRevB.99.205437},
  url = {https://link.aps.org/doi/10.1103/PhysRevB.99.205437}
}

@article{PhysRevA.97.022106,
  title = {Spin-chain model of a many-body quantum battery},
  author = {Le, Thao P. and Levinsen, Jesper and Modi, Kavan and Parish, Meera M. and Pollock, Felix A.},
  journal = {Phys. Rev. A},
  volume = {97},
  issue = {2},
  pages = {022106},
  numpages = {9},
  year = {2018},
  month = {Feb},
  publisher = {American Physical Society},
  doi = {10.1103/PhysRevA.97.022106},
  url = {https://link.aps.org/doi/10.1103/PhysRevA.97.022106}
}

@article{PhysRevLett.134.180401,
  title = {Topological Quantum Batteries},
  author = {Lu, Zhi-Guang and Tian, Guoqing and L\"u, Xin-You and Shang, Cheng},
  journal = {Phys. Rev. Lett.},
  volume = {134},
  issue = {18},
  pages = {180401},
  numpages = {8},
  year = {2025},
  month = {May},
  publisher = {American Physical Society},
  doi = {10.1103/PhysRevLett.134.180401},
  url = {https://link.aps.org/doi/10.1103/PhysRevLett.134.180401}
}

@article{PhysRevB.100.115142,
  title = {Many-body localized quantum batteries},
  author = {Rossini, Davide and Andolina, Gian Marcello and Polini, Marco},
  journal = {Phys. Rev. B},
  volume = {100},
  issue = {11},
  pages = {115142},
  numpages = {11},
  year = {2019},
  month = {Sep},
  publisher = {American Physical Society},
  doi = {10.1103/PhysRevB.100.115142},
  url = {https://link.aps.org/doi/10.1103/PhysRevB.100.115142}
}

@article{PhysRevLett.133.197001,
  title = {Controlling Energy Storage Crossing Quantum Phase Transitions in an Integrable Spin Quantum Battery},
  author = {Grazi, Riccardo and Sacco Shaikh, Daniel and Sassetti, Maura and Traverso Ziani, Niccol\'o and Ferraro, Dario},
  journal = {Phys. Rev. Lett.},
  volume = {133},
  issue = {19},
  pages = {197001},
  numpages = {6},
  year = {2024},
  month = {Nov},
  publisher = {American Physical Society},
  doi = {10.1103/PhysRevLett.133.197001},
  url = {https://link.aps.org/doi/10.1103/PhysRevLett.133.197001}
}

@article{GRAZI2025116383,
title = {Charging free fermion quantum batteries},
journal = {Chaos, Solitons {\&} Fractals},
volume = {196},
pages = {116383},
year = {2025},
issn = {0960-0779},
doi = {https://doi.org/10.1016/j.chaos.2025.116383},
url = {https://www.sciencedirect.com/science/article/pii/S0960077925003960},
author = {Riccardo Grazi and Fabio Cavaliere and Maura Sassetti and Dario Ferraro and Niccol{\`o} {Traverso Ziani}}
}

@Article{en18236116,
AUTHOR = {Grazi, Riccardo and Ferraro, Dario and Traverso Ziani, Niccol{\`o}},
TITLE = {Universal Features of Non-Analytical Energy Storage in Quantum Critical Quantum Batteries},
JOURNAL = {Energies},
VOLUME = {18},
YEAR = {2025},
NUMBER = {23},
ARTICLE-NUMBER = {6116},
URL = {https://www.mdpi.com/1996-1073/18/23/6116},
ISSN = {1996-1073},
DOI = {10.3390/en18236116}
}

@article{42rw-l31q,
  title = {Dynamical phase transitions and criticality-enhanced charging in a topological quantum battery},
  author = {Kheiri, Sasan and Kheiri, Siamak and Mahdavifar, Saeed},
  journal = {Phys. Rev. B},
  volume = {113},
  issue = {20},
  pages = {205150},
  numpages = {13},
  year = {2026},
  month = {May},
  publisher = {American Physical Society},
  doi = {10.1103/42rw-l31q},
  url = {https://link.aps.org/doi/10.1103/42rw-l31q}
}

@article{Heyl2013,
	author  = {M. Heyl and A. Polkovnikov and S. Kehrein},
	title   = {Dynamical Quantum Phase Transitions in the Transverse-Field Ising Model},
	journal = {Phys. Rev. Lett.},
	volume  = {110},
	pages   = {135704},
	year    = {2013},
	doi     = {10.1103/PhysRevLett.110.135704}
}

@article{PhysRevB.89.161105,
  title = {Disentangling dynamical phase transitions from equilibrium phase transitions},
  author = {Vajna, Szabolcs and D\'ora, Bal\'azs},
  journal = {Phys. Rev. B},
  volume = {89},
  issue = {16},
  pages = {161105},
  numpages = {5},
  year = {2014},
  month = {Apr},
  publisher = {American Physical Society},
  doi = {10.1103/PhysRevB.89.161105},
  url = {https://link.aps.org/doi/10.1103/PhysRevB.89.161105}
}

@article{PhysRevB.91.155127,
  title = {Topological classification of dynamical phase transitions},
  author = {Vajna, Szabolcs and D\'ora, Bal\'azs},
  journal = {Phys. Rev. B},
  volume = {91},
  issue = {15},
  pages = {155127},
  numpages = {5},
  year = {2015},
  month = {Apr},
  publisher = {American Physical Society},
  doi = {10.1103/PhysRevB.91.155127},
  url = {https://link.aps.org/doi/10.1103/PhysRevB.91.155127}
}

@article{PhysRevB.93.144306,
  title = {Slow quenches in a quantum Ising chain: Dynamical phase transitions and topology},
  author = {Sharma, Shraddha and Divakaran, Uma and Polkovnikov, Anatoli and Dutta, Amit},
  journal = {Phys. Rev. B},
  volume = {93},
  issue = {14},
  pages = {144306},
  numpages = {9},
  year = {2016},
  month = {Apr},
  publisher = {American Physical Society},
  doi = {10.1103/PhysRevB.93.144306},
  url = {https://link.aps.org/doi/10.1103/PhysRevB.93.144306}
}

@article{Heyl2018,
	author  = {M. Heyl},
	title   = {Dynamical quantum phase transitions: a review},
	journal = {Rep. Prog. Phys.},
	volume  = {81},
	pages   = {054001},
	year    = {2018},
	doi     = {10.1088/1361-6633/aaaf9a}
}

@article{PhysRevLett.121.130603,
  title = {Dynamical Quantum Phase Transitions: A Geometric Picture},
  author = {Lang, Johannes and Frank, Bernhard and Halimeh, Jad C.},
  journal = {Phys. Rev. Lett.},
  volume = {121},
  issue = {13},
  pages = {130603},
  numpages = {6},
  year = {2018},
  month = {Sep},
  publisher = {American Physical Society},
  doi = {10.1103/PhysRevLett.121.130603},
  url = {https://link.aps.org/doi/10.1103/PhysRevLett.121.130603}
}

@article{Heyl_2019,
doi = {10.1209/0295-5075/125/26001},
url = {https://doi.org/10.1209/0295-5075/125/26001},
year = {2019},
month = {feb},
publisher = {EDP Sciences, IOP Publishing and Societ{\`a} Italiana di Fisica},
volume = {125},
number = {2},
pages = {26001},
author = {Heyl, Markus},
title = {Dynamical quantum phase transitions: A brief survey},
journal = {Europhysics Letters}
}

@article{PhysRevB.100.085308,
  title = {Floquet dynamical quantum phase transitions},
  author = {Yang, Kai and Zhou, Longwen and Ma, Wenchao and Kong, Xi and Wang, Pengfei and Qin, Xi and Rong, Xing and Wang, Ya and Shi, Fazhan and Gong, Jiangbin and Du, Jiangfeng},
  journal = {Phys. Rev. B},
  volume = {100},
  issue = {8},
  pages = {085308},
  numpages = {11},
  year = {2019},
  month = {Aug},
  publisher = {American Physical Society},
  doi = {10.1103/PhysRevB.100.085308},
  url = {https://link.aps.org/doi/10.1103/PhysRevB.100.085308}
}

@article{PhysRevA.110.042209,
  title = {Exploring dynamical phase transitions in the $XY$ chain through a linear quench: Early and long-term perspectives},
  author = {Cao, Kaiyuan and Hou, Hongsheng and Tong, Peiqing},
  journal = {Phys. Rev. A},
  volume = {110},
  issue = {4},
  pages = {042209},
  numpages = {12},
  year = {2024},
  month = {Oct},
  publisher = {American Physical Society},
  doi = {10.1103/PhysRevA.110.042209},
  url = {https://link.aps.org/doi/10.1103/PhysRevA.110.042209}
}

@article{PhysRevB.110.064302,
  title = {Competition of long-range interactions and noise at a ramped quench dynamical quantum phase transition: The case of the long-range pairing Kitaev chain},
  author = {Baghran, R. and Jafari, R. and Langari, A.},
  journal = {Phys. Rev. B},
  volume = {110},
  issue = {6},
  pages = {064302},
  numpages = {11},
  year = {2024},
  month = {Aug},
  publisher = {American Physical Society},
  doi = {10.1103/PhysRevB.110.064302},
  url = {https://link.aps.org/doi/10.1103/PhysRevB.110.064302}
}

@article{srx7-cpl4,
  title = {Relation between equilibrium quantum phase transitions and dynamical quantum phase transitions in two-band systems},
  author = {Zeng, Yumeng and Chen, Shu},
  journal = {Phys. Rev. B},
  volume = {112},
  issue = {6},
  pages = {064307},
  numpages = {9},
  year = {2025},
  month = {Aug},
  publisher = {American Physical Society},
  doi = {10.1103/srx7-cpl4},
  url = {https://link.aps.org/doi/10.1103/srx7-cpl4}
}

@article{NIE2024130110,
title = {Flux-quench induced dynamical quantum phase transitions in an extended XY spin-chain},
journal = {Physica A: Statistical Mechanics and its Applications},
volume = {653},
pages = {130110},
year = {2024},
issn = {0378-4371},
doi = {https://doi.org/10.1016/j.physa.2024.130110},
url = {https://www.sciencedirect.com/science/article/pii/S0378437124006198},
author = {Wen-Hui Nie and Mei-Yu Zhang and Lin-Cheng Wang}
}

@article{PhysRevB.110.054312,
  title = {Entanglement in quenched extended Su-Schrieffer-Heeger model with anomalous dynamical quantum phase transitions},
  author = {Wong, Cheuk Yiu and Hui, Tsz Hin and Sacramento, P. D. and Yu, Wing Chi},
  journal = {Phys. Rev. B},
  volume = {110},
  issue = {5},
  pages = {054312},
  numpages = {17},
  year = {2024},
  month = {Aug},
  publisher = {American Physical Society},
  doi = {10.1103/PhysRevB.110.054312},
  url = {https://link.aps.org/doi/10.1103/PhysRevB.110.054312}
}

@article{Jurcevic2017,
  title = {Direct Observation of Dynamical Quantum Phase Transitions in an Interacting Many-Body System},
  author = {Jurcevic, P. and Shen, H. and Hauke, P. and Maier, C. and Brydges, T. and Hempel, C. and Lanyon, B. P. and Heyl, M. and Blatt, R. and Roos, C. F.},
  journal = {Phys. Rev. Lett.},
  volume = {119},
  issue = {8},
  pages = {080501},
  numpages = {5},
  year = {2017},
  month = {Aug},
  publisher = {American Physical Society},
  doi = {10.1103/PhysRevLett.119.080501},
  url = {https://link.aps.org/doi/10.1103/PhysRevLett.119.080501}
}

@article{Flaschner2018,
  author  = {Fl{\"a}schner, N. and Vogel, D. and Tarnowski, M. and Rem, B. S. and L{\"u}hmann, D.-S. and Heyl, M. and Budich, J. C. and Mathey, L. and Sengstock, K. and Weitenberg, C.},
  title   = {Observation of dynamical vortices after quenches in a system with topology},
  journal = {Nature Physics},
  year    = {2018},
  volume  = {14},
  number  = {3},
  pages   = {265--268},
  doi     = {10.1038/s41567-017-0013-8},
  url     = {https://doi.org/10.1038/s41567-017-0013-8},
  issn    = {1745-2481}
}

@article{Zhang2017,
  author  = {Zhang, J. and Pagano, G. and Hess, P. W. and Kyprianidis, A. and Becker, P. and Kaplan, H. and Gorshkov, A. V. and Gong, Z.-X. and Monroe, C.},
  title   = {Observation of a many-body dynamical phase transition with a 53-qubit quantum simulator},
  journal = {Nature},
  year    = {2017},
  volume  = {551},
  number  = {7682},
  pages   = {601--604},
  doi     = {10.1038/nature24654},
  url     = {https://doi.org/10.1038/nature24654},
  issn    = {1476-4687}
}

@article{Guo2019,
  title = {Observation of a Dynamical Quantum Phase Transition by a Superconducting Qubit Simulation},
  author = {Guo, Xue-Yi and Yang, Chao and Zeng, Yu and Peng, Yi and Li, He-Kang and Deng, Hui and Jin, Yi-Rong and Chen, Shu and Zheng, Dongning and Fan, Heng},
  journal = {Phys. Rev. Appl.},
  volume = {11},
  issue = {4},
  pages = {044080},
  numpages = {12},
  year = {2019},
  month = {Apr},
  publisher = {American Physical Society},
  doi = {10.1103/PhysRevApplied.11.044080},
  url = {https://link.aps.org/doi/10.1103/PhysRevApplied.11.044080}
}

@article{Gover2026,
  title = {Fully optimised variational simulation of a dynamical quantum phase transition on a trapped-ion quantum computer},
  author = {Gover, Lesley and Wimalaweera, Vinul and Azad, Fariha and DeCross, Matthew and Foss-Feig, Michael and Green, Andrew G.},
  journal = {Quantum},
  volume = {10},
  pages = {2109},
  year = {2026},
  month = {May},
  doi = {10.22331/q-2026-05-20-2109},
  url = {https://doi.org/10.22331/q-2026-05-20-2109}
}

@article{PhysRevLett.122.020501,
  title = {Simulating Dynamic Quantum Phase Transitions in Photonic Quantum Walks},
  author = {Wang, Kunkun and Qiu, Xingze and Xiao, Lei and Zhan, Xiang and Bian, Zhihao and Yi, Wei and Xue, Peng},
  journal = {Phys. Rev. Lett.},
  volume = {122},
  issue = {2},
  pages = {020501},
  numpages = {6},
  year = {2019},
  month = {Jan},
  publisher = {American Physical Society},
  doi = {10.1103/PhysRevLett.122.020501},
  url = {https://link.aps.org/doi/10.1103/PhysRevLett.122.020501}
}

@article{PhysRevLett.124.043001,
  title = {Observation of Dynamical Quantum Phase Transitions with Correspondence in an Excited State Phase Diagram},
  author = {Tian, T. and Yang, H.-X. and Qiu, L.-Y. and Liang, H.-Y. and Yang, Y.-B. and Xu, Y. and Duan, L.-M.},
  journal = {Phys. Rev. Lett.},
  volume = {124},
  issue = {4},
  pages = {043001},
  numpages = {6},
  year = {2020},
  month = {Jan},
  publisher = {American Physical Society},
  doi = {10.1103/PhysRevLett.124.043001},
  url = {https://link.aps.org/doi/10.1103/PhysRevLett.124.043001}
}

@article{Jordan1928,
  author  = {Jordan, P. and Wigner, E.},
  title   = {{\"U}ber das Paulische {\"A}quivalenzverbot},
  journal = {Zeitschrift f{\"u}r Physik},
  year    = {1928},
  volume  = {47},
  number  = {9},
  pages   = {631--651},
  doi     = {10.1007/BF01331938},
  url     = {https://doi.org/10.1007/BF01331938},
  issn    = {0044-3328}
}

@article{PhysRevA.2.1075,
  title = {Statistical Mechanics of the $\mathrm{XY}$ Model. I},
  author = {Barouch, Eytan and McCoy, Barry M. and Dresden, Max},
  journal = {Phys. Rev. A},
  volume = {2},
  issue = {3},
  pages = {1075--1092},
  numpages = {0},
  year = {1970},
  month = {Sep},
  publisher = {American Physical Society},
  doi = {10.1103/PhysRevA.2.1075},
  url = {https://link.aps.org/doi/10.1103/PhysRevA.2.1075}
}

@article{PhysRevLett.97.200601,
  title = {Sweeping from the Superfluid to the Mott Phase in the Bose-Hubbard Model},
  author = {Sch\"utzhold, Ralf and Uhlmann, Michael and Xu, Yan and Fischer, Uwe R.},
  journal = {Phys. Rev. Lett.},
  volume = {97},
  issue = {20},
  pages = {200601},
  numpages = {4},
  year = {2006},
  month = {Nov},
  publisher = {American Physical Society},
  doi = {10.1103/PhysRevLett.97.200601},
  url = {https://link.aps.org/doi/10.1103/PhysRevLett.97.200601}
}

@article{PhysRevA.77.043615,
  title = {Bogoliubov theory of quantum correlations in the time-dependent Bose-Hubbard model},
  author = {Fischer, Uwe R. and Sch\"utzhold, Ralf and Uhlmann, Michael},
  journal = {Phys. Rev. A},
  volume = {77},
  issue = {4},
  pages = {043615},
  numpages = {10},
  year = {2008},
  month = {Apr},
  publisher = {American Physical Society},
  doi = {10.1103/PhysRevA.77.043615},
  url = {https://link.aps.org/doi/10.1103/PhysRevA.77.043615}
}

@article{PhysRevLett.99.120407,
  title = {Vortex Quantum Creation and Winding Number Scaling in a Quenched Spinor Bose Gas},
  author = {Uhlmann, Michael and Sch\"utzhold, Ralf and Fischer, Uwe R.},
  journal = {Phys. Rev. Lett.},
  volume = {99},
  issue = {12},
  pages = {120407},
  numpages = {4},
  year = {2007},
  month = {Sep},
  publisher = {American Physical Society},
  doi = {10.1103/PhysRevLett.99.120407},
  url = {https://link.aps.org/doi/10.1103/PhysRevLett.99.120407}
}

@article{PhysRevD.81.025017,
  title = {$O(N)$ symmetry-breaking quantum quench: Topological defects versus quasiparticles},
  author = {Uhlmann, Michael and Sch\"utzhold, Ralf and Fischer, Uwe R.},
  journal = {Phys. Rev. D},
  volume = {81},
  issue = {2},
  pages = {025017},
  numpages = {6},
  year = {2010},
  month = {Jan},
  publisher = {American Physical Society},
  doi = {10.1103/PhysRevD.81.025017},
  url = {https://link.aps.org/doi/10.1103/PhysRevD.81.025017}
}

@article{Uhlmann_2010,
doi = {10.1088/1367-2630/12/9/095020},
url = {https://doi.org/10.1088/1367-2630/12/9/095020},
year = {2010},
month = {sep},
publisher = {},
volume = {12},
number = {9},
pages = {095020},
author = {Uhlmann, Michael and Sch{\"u}tzhold, Ralf and R Fischer, Uwe},
title = {System size scaling of topological defect creation in a second-order dynamical quantum phase transition},
journal = {New Journal of Physics}
}

@article{10.1116/5.0184903,
    author = {Gyhm, Ju-Yeon and Fischer, Uwe R.},
    title = {Beneficial and detrimental entanglement for quantum battery charging},
    journal = {AVS Quantum Science},
    volume = {6},
    number = {1},
    pages = {012001},
    year = {2024},
    month = {01},
    issn = {2639-0213},
    doi = {10.1116/5.0184903},
    url = {https://doi.org/10.1116/5.0184903},
}

@article{SSH,
  author = {Su, W. P. and Schrieffer, J. R. and Heeger, A. J.},
  title = {Solitons in Polyacetylene},
  journal = {Physical Review Letters},
  volume = {42},
  number = {25},
  pages = {1698--1701},
  year = {1979},
  doi = {10.1103/PhysRevLett.42.1698}
}

@article{Selke,
  author = {Selke, Walter},
  title = {The ANNNI model---Theoretical analysis and experimental application},
  journal = {Physics Reports},
  volume = {170},
  number = {4},
  pages = {213--264},
  year = {1988},
  doi = {10.1016/0370-1573(88)90140-8}
}

@article{LiebRobinson,
  author = {Lieb, Elliott H. and Robinson, Derek W.},
  title = {The finite group velocity of quantum spin systems},
  journal = {Communications in Mathematical Physics},
  volume = {28},
  pages = {251--257},
  year = {1972},
  doi = {10.1007/BF01645779}
}

@article{ZanardiPaunkovic2006,
  author = {Zanardi, Paolo and Paunkovi{\'c}, Nikola},
  title = {Ground state overlap and quantum phase transitions},
  journal = {Phys. Rev. E},
  volume = {74},
  pages = {031123},
  year = {2006},
  doi = {10.1103/PhysRevE.74.031123}
}

@article{YouLiGu2007,
  author = {You, Wen-Long and Li, Ying-Wai and Gu, Shi-Jian},
  title = {Fidelity, dynamic structure factor, and susceptibility in critical phenomena},
  journal = {Phys. Rev. E},
  volume = {76},
  pages = {022101},
  year = {2007},
  doi = {10.1103/PhysRevE.76.022101}
}

@article{Gu2010Fidelity,
  author = {Gu, Shi-Jian},
  title = {Fidelity approach to quantum phase transitions},
  journal = {Int. J. Mod. Phys. B},
  volume = {24},
  pages = {4371--4458},
  year = {2010},
  doi = {10.1142/S0217979210056335}
}
